\documentclass{aa}
\usepackage[utf8]{inputenc}
\usepackage{natbib}
\usepackage{graphicx}
\usepackage[varg]{txfonts}
\usepackage{amssymb}

\newcommand{\Lp}{L$^{\prime}$~}
\newcommand{\MJ}{$M_{\textrm{Jup}}$~}

\bibpunct{(}{)}{;}{a}{}{,}

\begin{document}

   \title{Combining high contrast imaging and radial velocities to constrain the planetary architecture of nearby stars\thanks{ Based on observations collected at the European Southern Observatory, Chile ESO No.~082.C-0518 and 084.C-0656.}}

   \author{A. Boehle\inst{1}\fnmsep\thanks{National Center of Competence in Research “PlanetS” (\protect\url{https://nccr-planets.ch}).}, S.~P. Quanz\inst{1}, C. Lovis\inst{2}, D. S\'{e}gransan\inst{2}, S. Udry\inst{2}, \and D. Apai\inst{3}
          }

   \institute{Institute for Particle Physics and Astrophysics, ETH Zurich, 8093 Zurich, Switzerland \\
             \email{anna.boehle@phys.ethz.ch}
             \and
             Observatoire Astronomique de l’Universit\'{e} de Gen\`{e}ve, 51 Ch. des Maillettes, 1290 Versoix, Switzerland
             \and
             Department of Astronomy/Steward Observatory, The University of Arizona, 933 North Cherry Avenue, Tucson, AZ 85721, USA
             }

    \authorrunning{A. Boehle, et al.}
   \date{Received April 19, 2019; accepted July 2, 2019}

 \abstract
   {Nearby stars are prime targets for exoplanet searches and characterization using a variety of detection techniques. Combining constraints from the complementary detection methods of high contrast imaging (HCI) and radial velocity (RV) can further constrain the planetary architectures of these systems because these methods place limits at different regions of the companion mass and semi-major axis parameter space. Compiling a census of the planet population in the solar neighborhood is important to inform target lists for future space missions that will specifically target nearby stars to search for Earth analogues.}
   {We aim to constrain the planetary architectures from the combination of HCI and RV data for 6 nearby stars within 6 pc: $\tau$ Ceti, Kapteyn's star, AX Mic, 40 Eri, HD 36395, and HD 42581. We explored where HCI adds information to constraints from the long-term RV monitoring data for these stars.}
   {We compiled the sample from stars with available archival VLT/NACO HCI data at \Lp band (3.8 $\mu$m), where we expect substellar companions to be brighter for the typically older ages of nearby field stars (>1 Gyr). The NACO data were fully reanalyzed using the state-of-the-art direct imaging pipeline PynPoint and combined with RV data from HARPS, Keck/HIRES, and CORALIE. A Monte Carlo approach was used to assess the completeness in the companion mass/semi-major axis parameter space from the combination of the HCI and RV data sets.}
   {We find that the HCI data add significant information to the RV constraints, increasing the completeness for certain companions masses/semi-major axes by up to 68 -- 99\% for 4 of the 6 stars in our sample, and by up to 1 -- 13\% for the remaining stars. The improvements are strongest for intermediate semi-major axes (15 –- 40 AU), corresponding to the semi-major axes of the ice giants in our own solar system.  The HCI mass limits reach 5  –-  20 \MJ in the background-limited regime, depending on the age of the star.}
   {Through the combination of HCI and RV data, we find that stringent constraints can be placed on the possible substellar companions in these systems. Applying these methods systematically to nearby stars will quantify our current knowledge of the planet population in the solar neighborhood and inform future observations.}


\maketitle

\section{Introduction} \label{sec:intro}

 Nearby stars (<20 pc) are particularly interesting targets for exoplanet searches and characterization using a variety of techniques. These stars are bright enough for precise radial velocity (RV) measurements, and many have been the targets of long-term RV monitoring with high-resolution spectrographs such as HARPS \citep{2002Msngr.110....9P,2003Msngr.114...20M} and Keck/HIRES \citep{1994SPIE.2198..362V}. 
 The proximity of these stars also improves the companion masses that can be detected through direct or high contrast imaging (HCI) in the background-limited regime by increasing the apparent magnitude of their thermal emission \citep[e.g.,][]{Thalmann_2011, 2012ApJ...754..127Q, Mawet_2019}. 
 The current space missions TESS \citep{2015JATIS...1a4003R} and Gaia \citep[see][]{2014ApJ...797...14P} are specifically focused on searches for planets around nearby stars using the transit and astrometry techniques, respectively. Future large-aperture missions now under study, such as HabEx and LUVOIR, or a space-based mid-infrared interferometer will also target nearby stars to directly image and characterize many Earth-like planets \citep{2016SPIE.9904E..0LM,2017SPIE10398E..0FP,2018A&A...609A...4K}. Constraining the planet population in the solar neighborhood using current data and facilities will help prioritize targets for these missions as well as future ground-based instruments and telescopes such as VLT/ERIS \citep{2018SPIE10702E..09D}, ELT/METIS \citep{2015IJAsB..14..279Q,2018SPIE10702E..1UB}, and ELT/HiRES \citep{2018SPIE10702E..1YM}.

To quantify our current constraints on the planetary architecture of nearby stars, it is advantageous to combine the various exoplanet detection methods since these probe complimentary parameter space 
 in companion mass and semi-major axis 
 and can give information on different exoplanet parameters. The RV method is most sensitive to close-in planets and yields estimates of the planet's minimum mass and some of its orbital parameters, while HCI is most sensitive to massive planets at wide separations and measures photons from the planet itself that can be used to characterize the planet's atmosphere and estimate its mass. When a planet is detected using both RV and HCI, and more recently by adding information from \textit{Hipparcos} and \textit{Gaia} Data Release 2, it is possible to break the planet mass/inclination degeneracy from RV data alone 
 and the resulting dynamical mass can then be compared to the masses derived from evolutionary and atmospheric models to further refine these predictions \citep[e.g.,][]{2012ApJ...751...97C, 2018AJ....155..159B, 2018A&A...614A..16C, 2018NatAs...2..883S,2018arXiv181107285B, Dupuy_2019}. 
Even for cases in which no planet is detected, the complimentary nature of these different detection methods can put substantial constraints on the possible masses and semi-major axes of substellar companions in individual systems.

In this work, we focus on combining constraints from HCI and RV data to build a census of planets in the solar neighborhood. The sample presented here consists of 6 stars that are all within 6 pc, which have been targets of long-term RV monitoring campaigns and have available archival HCI data.  We specifically consider stars with available HCI data at the \Lp band ($\lambda_0$ = 3.8 $\mu$m) from the NACO instrument at the VLT \citep{2003SPIE.4839..140R,2003SPIE.4841..944L}. Cool substellar companions are expected to be brighter at these thermal infrared (IR) wavelengths than at shorter wavelengths and thus these data can reach better mass constraints even with weaker contrast limits compared to extreme adaptive optics (AO) near-infrared instruments such as VLT/SPHERE and Gemini/GPI \citep[e.g.,][]{2010ApJ...714.1551H, 2015MNRAS.454..129V, 2019arXiv190204080B, 2014PNAS..11112661M}. 
Here we consider the constraints on a system's planetary architecture that can be derived from the combination of HCI and RV data when no massive companion is detected, in contrast to the works mentioned above that combined HCI and RV data to measure dynamical masses of known companions.

This paper is organized as follows. In Sec.~\ref{sec:obs}, the archival HCI data and the RV data are described. In Sec.~\ref{sec:analysis}, we derive mass limits from the HCI data and combine these with companion mass/semi-major axis constraints from the RV data. The resulting limits on the planetary architecture of these systems are presented in Sec.~\ref{sec:results} and discussed in Sec.~\ref{sec:disc}.

\section{Sample and data reduction} \label{sec:obs}

\subsection{High contrast imaging data}
\label{sec:hcidata}

\begin{table*}
\caption{Stellar properties\label{tab:stellar_prop} }
\centering
\begin{tabular}{llccccr}
  \hline \hline 
 Star & Spectral type & Distance & Apparent magnitude &  Age [min. - max.] &  Mass &  References \\ 
  & & (pc) & ($L^{\prime}$) & (Gyr) & (M$_{\odot}$) &  \\ 
 \hline 
$\tau$ Ceti & G8V & 3.65 & 1.7 & 5.8 [2.9 - 8.7] & 0.63 & 1, 2, 3, 4, 5 \\ 
Kapteyn's star & M1VIp & 3.93 & 4.9 & 1.7 [0.5 - 2.9] & 0.27 & 5, 6, 7, 8, 9 \\ 
AX Mic & M1V & 3.97 & 2.9 & 4.8 [1.9 - 7.7] & 0.56 & 5, 6, 8, 10, 11, 12 \\ 
40 Eri & K0V & 4.98 & 2.4 & 5.6 [2.8 - 8.4] & 0.67 & 1, 3, 4, 5, 10 \\ 
HD 36395 & M1.5Ve & 5.70 & 3.8 & 5.8 [2.7 - 8.8] & 0.64 & 5, 6, 7, 9, 13 \\ 
HD 42581 & M1V & 5.76 & 4.0 & 1.6 [0.3 - 3.0] & 0.58 & 6, 7, 9, 14, 15 \\ 
\hline 
\end{tabular} 
\tablefoot{ 
References: (1) \citet{2007A&A...474..653V} - distance, (2) \citet{2003AJ....125.3311K} - mag., (3) \citet{2008ApJ...687.1264M} - age, (4) \citet{2008A&A...487..373S} - mass, (5) \citet{1989ApJS...71..245K} - spectral type, (6) \citet{2018A&A...616A...1G} - distance, (7) \citet{2013yCat.2328....0C} - mag., (8) \citet{2013ApJ...768...25G} - age, (9) \citet{2017A&A...600A..13A} - mass, (10) \citet{1978A&AS...34..477M} - mag., (11) \citet{2012AJ....143..135V} - age, (12) \citet{2007A&A...468..663T} - mass, (13) \citet{2017ApJ...836...77Y} - age, (14) \citet{2015AJ....150...53N} - age, (15) \citet{1991ApJS...77..417K} - spectral type. 
} 
\end{table*}

\begin{table*}
\caption{Summary of Archival NACO Imaging Observations\label{tab:NACO_obs} }
\centering
\begin{tabular}{lcccccccc}
  \hline \hline 
 Star & Date & $t_{\text{int}}$/frame &   \# of frames & \# of frames & Total $t_{\text{int}}$ & Airmass &   Field rotation &   Program ID \\ 
  & (UT) & (sec) & used & removed & (min) &  & (deg) & \\ 
 \hline 
$\tau$ Ceti & 2008-11-16 & 0.175 & 4137 & 3 & 12 & 1.02 - 1.07 & 19 & 082.C-0518 \\ 
Kapteyn's star & 2008-11-16 & 0.500 & 2815 & 20 & 23 & 1.10 - 1.14 & 15 & 082.C-0518 \\ 
AX Mic & 2009-10-11 & 0.175 & 3423 & 10 & 10 & 1.07 - 1.12 & 14 & 084.C-0656 \\ 
40 Eri & 2008-11-16 & 0.500 & 1268 & 1693 & 11 & 1.05 - 1.07 & 21 & 082.C-0518 \\ 
HD 36395 & 2008-11-16 & 0.300 & 3842 & 1 & 19 & 1.08 - 1.12 & 18 & 082.C-0518 \\ 
HD 42581 & 2008-11-16 & 0.300 & 3723 & 120 & 19 & 1.00 - 1.01 & 53 & 082.C-0518 \\ 
\hline 
\end{tabular} 
\end{table*}

The sample of stars for this work was constructed by searching the ESO archive for programs targeting very nearby stars ($<6$ pc) with the NACO instrument at the VLT. NACO consists of NAOS, an AO system, and CONICA, a 1 -- 5 $\mu$m imager and spectrograph. As discussed above, we selected stars with data taken in the \Lp filter ($\lambda_0$ = 3.80 $\mu$m, $\Delta\lambda$ = 0.62  $\mu$m), since planets are expected to be brighter at these thermal infrared wavelengths for the old ages ($>1$ Gyr) of solar neighborhood stars. It was also required that the data were taken in pupil tracking mode to enable angular differential imaging \citep[ADI;][]{2006ApJ...641..556M} and in ``cube'' mode, so that each individual short (0.175 -- 0.5 sec) exposure is written out. 
The properties of the resulting sample of 6 stars is detailed in Table \ref{tab:stellar_prop}. 

The NACO HCI data for all 6 stars in the sample are summarized in Table \ref{tab:NACO_obs}. 
The data primarily come from the ESO 082.C-05182 program (PI: Apai), which targeted all sources within 6 pc and RV planet hosts within 13 pc. The data on the star AX Mic are from the ESO 084.C-0656 program (PI: Apai), which targeted the HR 8799 planets. All the archival data analyzed in this work have not previously been published. The data were taken with the AO system using the target star as the natural guide star and without a coronagraph. These observations also employed a dither pattern that moves the star from the center of one detector quadrant to another every few data cubes to facilitate the subtraction of the strong sky background at \Lp. Due to the lack of coronagraph and the brightness of these nearby stars, all data are saturated in the stellar core.

\begin{figure*}[t]
	\begin{center}
	\includegraphics[width=\textwidth]{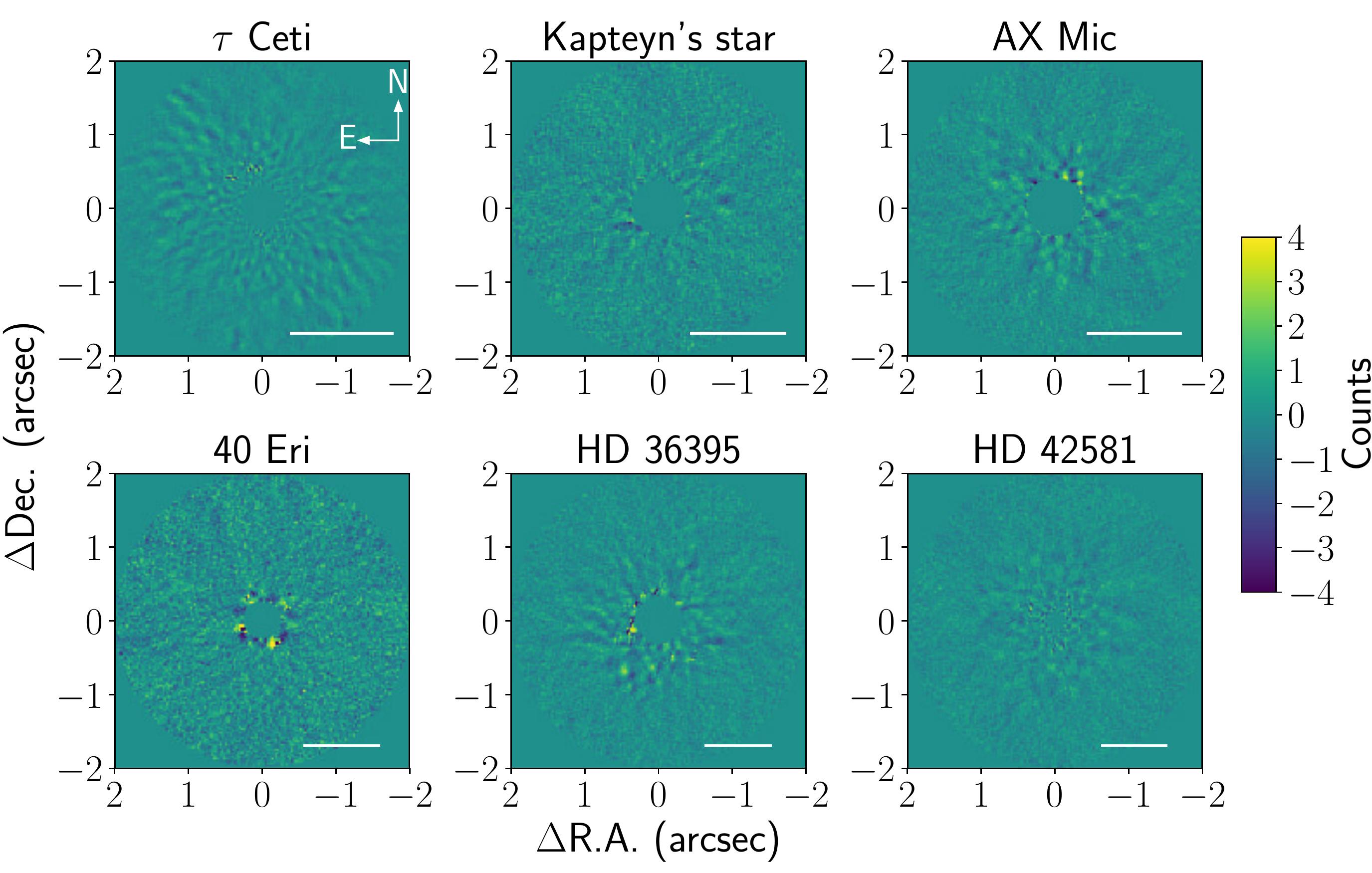}  
	\end{center}
	\caption
	{ PSF-subtracted images for each star in the sample, where the PSF was modeled using Principal Component Analysis. These images were used to search for substellar companions with projected separations out to 2\arcsec~from the star. The central masked region was determined by the amount of field rotation in the data. The white bar in the lower left of each image corresponds to 5 AU at the distance of the star. The images are plotted with a linear stretch.
	\label{fig:imgs_pca}
    }
\end{figure*}

The raw NACO HCI data were downloaded from the ESO archive and reduced using the state-of-the-art direct imaging data pipeline PynPoint \citep[][v0.5.3]{2012MNRAS.427..948A,2019A&A...621A..59S}.  PynPoint includes modules for calibration steps as well as modules to subtract the stellar point spread function (PSF) using Principal Component Analysis (PCA).  All data were dark, flat, and bad pixel corrected, and then the sky was subtracted using an average of the preceding and subsequent data cubes 
in which the star was at a different dither position.  
Automatic frame selection was performed to remove any frames in which the AO loops were open. This frame selection was done by performing aperture photometery of the star in each frame using a circular aperture with a radius of 0\farcs2, and then removing frames in which the measured stellar counts differed from the median by more than 4 times the standard deviation. 
Manual frame selection was also performed for 40 Eri because these data suffered from a tracking issue that caused the star to be close to the edge of the detector in over half of the data set. Next, the star was located in each frame and an area of 4\arcsec~by 4\arcsec~around the star was cropped out. The resulting images were aligned relative to each other using cross correlation of the full 4\arcsec~by 4\arcsec~images. The aligned images were then centered by fitting a Gaussian to a central circular region of 1\arcsec~in radius in the average of all images.  The cross correlation and the Gaussian fit were performed over large regions of the images so that the alignment and centering are accurate despite the core of the stellar PSF being saturated.  Finally, the stellar PSF was modeled using PCA after masking out a central region corresponding to minimum separation at which a planet could be detected as determined by the field rotation of the data (see Sec.~\ref{sec:hci_lims} for details). The final reduced and PSF-subtracted image for each star is shown in Fig.~\ref{fig:imgs_pca}. 

\begin{figure*}[t]
	\begin{center}
	\includegraphics[width=\textwidth]{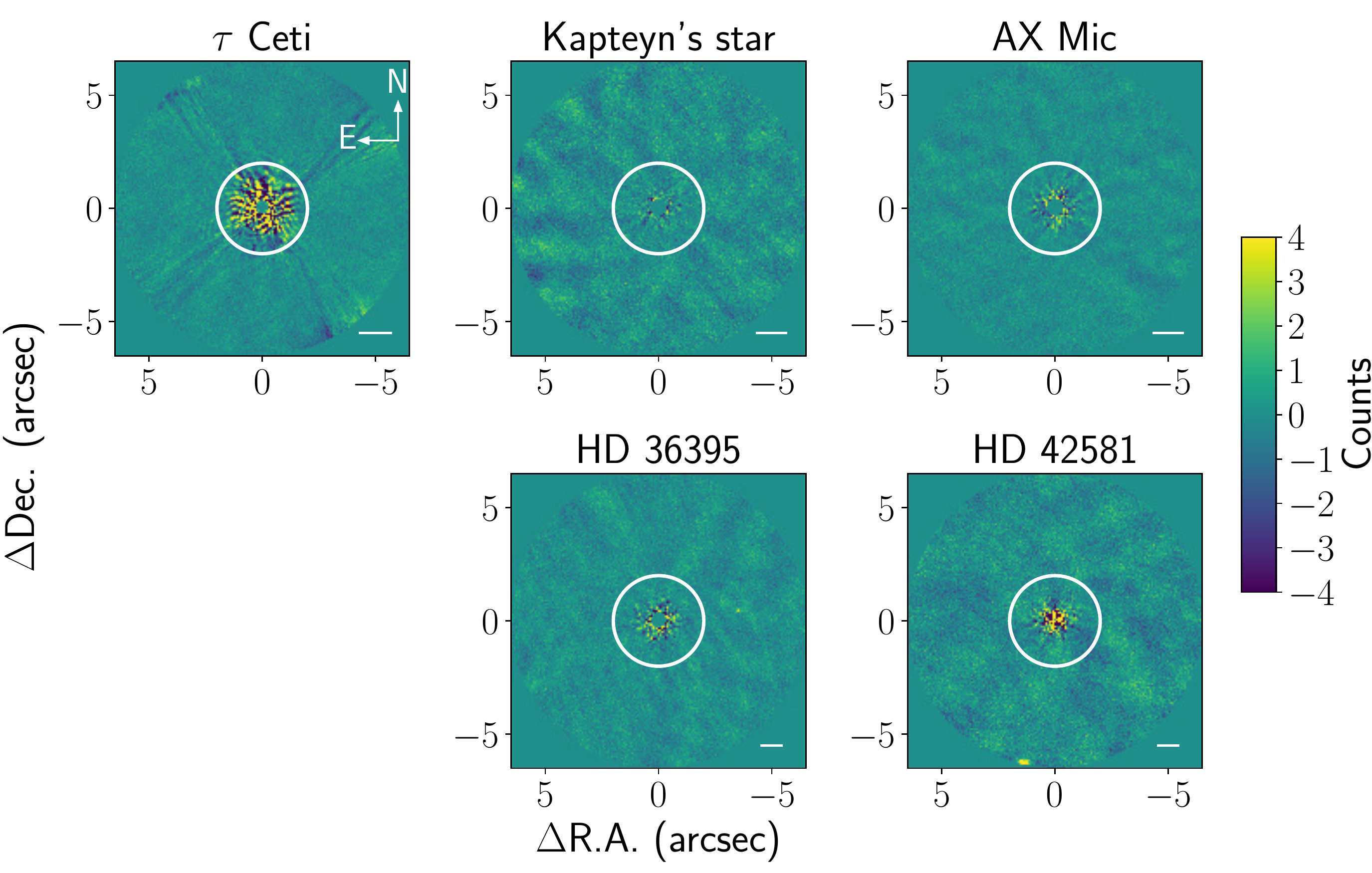} 
	\end{center}
	\caption
	{ PSF-subtracted images for each star in the sample, where the PSF was subtracted using classical ADI. These images were used to search for companions from 2\arcsec~out to 6.5\arcsec~(the maximum separation allowed by the NACO field of view and the dithering observing strategy). The white circle indicates the extent of the PCA images shown in Fig.~\ref{fig:imgs_pca} (2\arcsec~radius). The white bar in the lower left of each image corresponds to 5 AU at the distance of the star. The images are plotted with a linear stretch.
	\label{fig:imgs_adi}
    }
\end{figure*}

To search for substellar companions at projected separations outside of 2\arcsec, classical ADI was performed \citep[specifically, the mean of all images was subtracted from each image before de-rotating and averaging;][]{2006ApJ...641..556M}. At these larger separations from the star, the subtraction of the stellar PSF is less critical as the data are in the background-limited regime and no longer the contrast-limited regime. Prior to the ADI PSF subtraction, the individual frames were cropped to 13\arcsec~by 13\arcsec~and the same central region as above was masked out. This 13\arcsec~field of view (FOV) is set by the size of one quadrant of the NACO detector at the plate scale of $\sim$ 27 mas. 
The classical ADI subtraction was not performed for 40 Eri because the star was too close to the edge of the detector to probe separations larger than 2\arcsec~for all position angles. The final classical ADI images are shown in Fig.~\ref{fig:imgs_adi}.

\subsection{RV data}
\label{sec:rvdata}

RV data were taken from the Data Analysis Center for Exoplanets\footnote{\url{https://dace.unige.ch/} (site accessed on 11 January 2019).} (DACE) hosted by the University of Geneva, which compiles available RV measurements from a variety of sources. The RV data for the targets in this work consist of new and previously published HARPS data (for $\tau$ Ceti, 40 Eri, HD 42581, and Kapteyn's Star), published Keck/HIRES data \citep[][for all stars]{2017AJ....153..208B}, and CORALIE data (for $\tau$ Ceti). CORAVEL data for $\tau$ Ceti are also available in the DACE database, but these data were removed due to their significantly higher scatter compared to the other data sets (270 m/s versus 2.8 m/s). 
 
Using the analysis tools available on DACE, the RV data were binned on a nightly basis. A nightly binning was chosen because this work focuses on companions with orbital periods much longer than a 1 day timescale. The binned RV time series were then downloaded directly from DACE and then the secular acceleration of each star was corrected given the star's parallax and proper motion from the parallax reference listed in Table \ref{tab:stellar_prop}. 
The RV offsets of the different instruments were then corrected. For this step, each data release from each instrument (e.g., COR98, COR07, and COR14 from the CORALIE instrument) was treated independently, since upgrades or other changes to the instrument can introduce different RV offsets. 
Each RV offset was determined by adding the data sets one by one from the earliest to the latest starting time and setting the mean of the temporarily overlapping data points to 0. 
The HARPS15 data sets for Kapteyn's star and HD 42581 do not overlap in time with any other RV data set, so for these stars, the HARPS15 data were instead anchored to the HARPS03 data using an average of the offsets measured for the two RV standard stars with an early M spectral type (M2.5V) similar to Kapteyn's star and HD 42581 in \citet{2015Msngr.162....9L}. For these HARPS15 data, the dispersion of the measured RV offsets for the standard stars (2 m/s) is added in quadrature to the HARPS15 error. We note that the uncertainty in the HARPS15 RV offset is likely even larger for a given star, since the value depends on both the spectral type and the $v~\textrm{sin}~i$ of the star. 
Finally, the overall standard deviation of the calibrated RV data was taken for use in the subsequent analysis.

\begin{figure*}[t]
	\begin{center}
	\includegraphics[width=\textwidth]{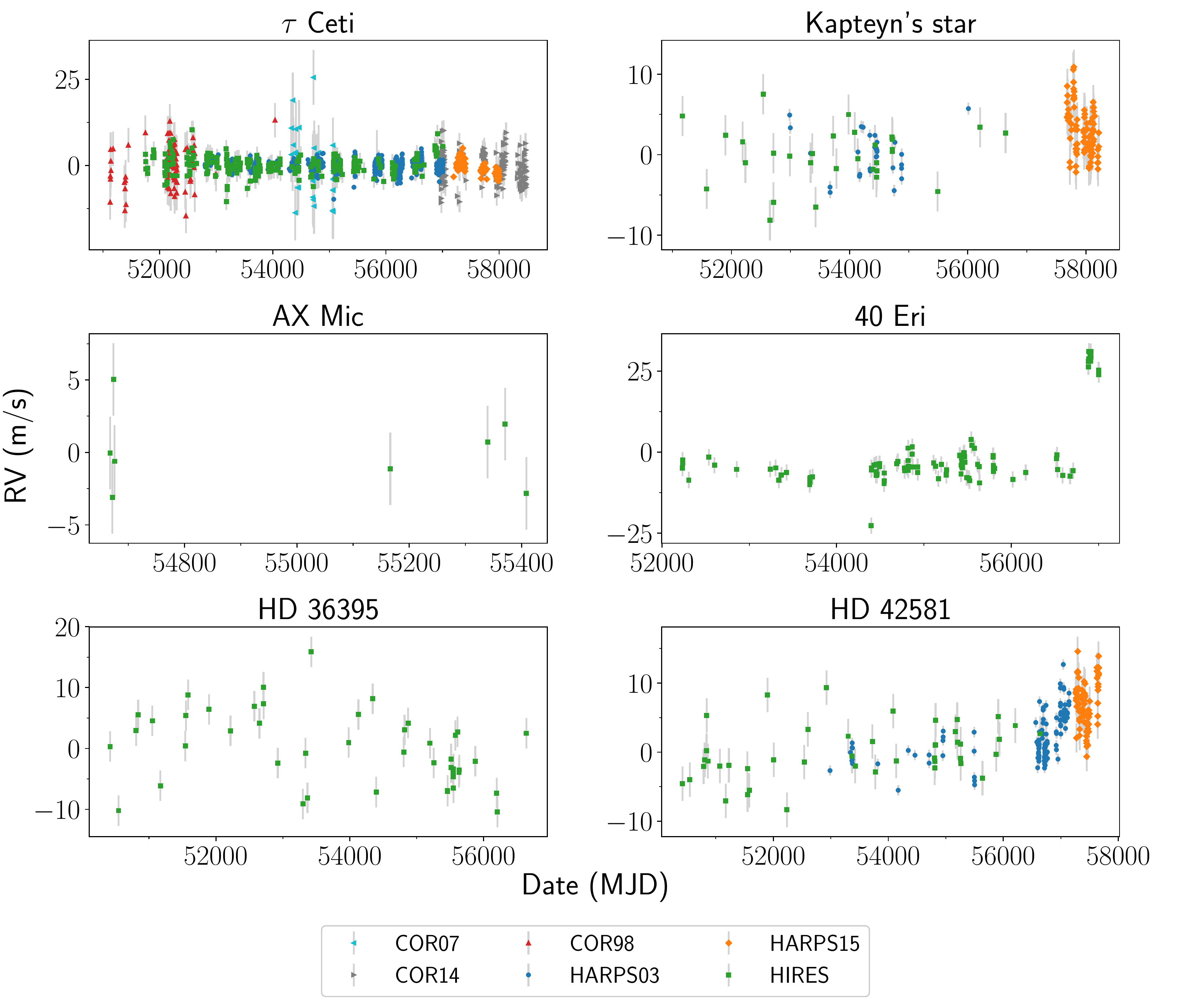}  
	\end{center}
	\caption
	{Radial velocity data for each star in the sample. Data from CORALIE (COR98, COR07, COR14), HARPS (HARPS03, HARPS15) and HIRES are shown, with each data set plotted as a separate marker/color.
	\label{fig:rv_data}
    }
\end{figure*}

The resulting calibrated RV data are shown in Fig.~\ref{fig:rv_data} and Table \ref{tab:RV_obs} summarizes the properties of these RV data for each target star. The time baseline of the RV observations ranges from just over 2 years for AX Mic to $\sim$20 years for $\tau$ Ceti. The scatter in the RV measurements also spans a wide range from 2.8 to 12.4 m/s. The larger RV scatter values are likely due to stellar activity. Instrumental effects could also play a roll in the case of 40 Eri, which exhibits a jump in the measured RV in the most recent HIRES epochs that is uncorrelated with activity index changes and that drives this star's high RV scatter. 

\begin{table*}
\caption{Summary of RV Observations\label{tab:RV_obs} }
\centering
\begin{tabular}{lccccccc}
  \hline \hline 
 Star & \multicolumn{4}{c}{\# of measurements} & $\Delta$t$_{\textrm{RV}}\tablefootmark{a}$ & $a$ for $\Delta$t$_{\textrm{RV}}\tablefootmark{b}$ & $\sigma_{\textrm{RV}}$  \\ 
  & HARPS & HIRES & CORALIE & total  & (days) & (AU) & (m/s)  \\ 
 \hline 
$\tau$ Ceti & 1038 & 272 & 314 & 1624 & 7353 & 6.3 & 2.8 \\ 
Kapteyn's star & 125 & 31 & 0 & 156 & 7046 & 4.7 & 3.3 \\ 
AX Mic & 0 & 8 & 0 & 8 & 740 & 1.3 & 2.6 \\ 
40 Eri & 0 & 109 & 0 & 109 & 4766 & 4.9 & 12.4 \\ 
HD 36395 & 0 & 41 & 0 & 41 & 6220 & 5.7 & 6.0 \\ 
HD 42581 & 198 & 44 & 0 & 242 & 7240 & 6.1 & 4.4 \\ 
\hline 
\end{tabular} 
\tablefoot{ 
\tablefoottext{a}{Total time baseline of RV observations.} \\ 
\tablefoottext{b}{Semi-major axis corresponding to an orbital period that equals the time baseline of the RV observations given the stellar mass in Table \ref{tab:stellar_prop}.} \\ 
}
\end{table*}

\section{Analysis}
\label{sec:analysis}

\subsection{High contrast imaging contrast and mass limits}
\label{sec:hci_lims}
We first searched the PSF-subtracted images for any companion candidates. A point source is detected in the HD 42581 data to the SSE whose position is consistent with the known brown dwarf companion HD 42581 B \citep{1995Natur.378..463N,1995Sci...270.1478O}. 
We also identified a point source to the west of HD 36395 with a separation of 3.5\arcsec, a PA of 276 degrees, and a S/N of 16.5 \citep[where S/N is defined as in][]{2014ApJ...792...97M}. 
We searched for possible background sources coinciding with this candidate in a range of published catalogues and found that the position of this detection is consistent with that of a faint optical source (B mag = 19.9) in the Guide Star Catalog V2.3 within the reported errors \citep[GSC2.3 identifier: S1LI000310;][]{2008AJ....136..735L}. 
The position is also consistent with a source in the GAIA DR2 catalogue \citep[BP = 19.2 mag, GAIA DR2 identifier: 3209938267881341056;][]{2018A&A...616A...1G}. 
This GAIA source has a measured parallax of $0.7\pm0.4$ mas, which corresponds to a distance of 0.9 -- 3 kpc. Based on these catalogue identifications, we conclude that this candidate companion is a background star. 
Finally, a second possible point source to the SE of HD 36395 is visible at a separation of 0.84\arcsec and a PA of 145 degrees, but this detection is not significant (S/N = 3.2). Having found no new significant point sources in the data, we focus the rest of the analysis and discussion on the HCI detection limits.

Contrast limits for the final HCI images were derived by planting artificial companions in the individual frames, performing the full PCA or ADI PSF subtraction, and then computing the resulting detection significance.  This method accounts for the self-subtraction of the planet signal that occurs when the stellar PSF is modeled using PCA or subtracted using ADI.  
We tested a range of values for the number of principal components (PCs) and the number of images that were temporally stacked prior to PSF subtraction to determine the optimal parameters for each data set.  For these tests, the limiting contrast was computed at 4 evenly-spaced separations from the edge of the central mask (0\farcs1 -- 0\farcs4; see Table \ref{tab:psf_params}) out to the edge of the 4 by 4\arcsec~field of view and along a position angle of 0 deg for a number of PCs ranging from 0.5 to 40\% of the total frames and a number of stacked frames from 1 to 40.  For each stacking value, the number of PCs that yielded the two best contrast limits were then selected and tested at 3 additional position angles.  Finally, the optimal combination of stacking and number of PCs was chosen by prioritizing the wider 2 separations, since these separations are more likely to contribute additional constraints to those from the RV data alone. We note that by optimizing for the contrast at wide separations, the closest separations can have substantial self-subtraction and in some cases a detection with the specified significance could not be achieved. 
For the classical ADI PSF subtraction, we also tested the same range of stacking values and found that changing the number of stacked frames did not affect the final contrast limits. In the case of HD 42581, we masked out the region around HD 42581 B when determining the noise at the widest separation. The final parameters for each star used in the PSF subtraction are presented in Table \ref{tab:psf_params}.

The final contrast curves with the optimal PSF subtraction parameters were computed at steps of 0\farcs1 (a total of 40 -- 60 separations for stars other than 40 Eri) by averaging over the contrasts at 3 positions angles (0, 120, and 240 degrees). We used an average of 3 position angles due to constraints on computational time: determining the contrast for each image position required running the full PSF subtraction multiple times to find the contrast with the specified significance.  
We computed the contrast for a constant false positive fraction (FPF) of $2.9 \times 10^{-7}$, while accounting for the small sample statistics at close separations \citep{2014ApJ...792...97M}. This FPF is equivalent to a 5-sigma detection threshold for an underlying Gaussian distribution and a large number of samples.  We note that contrast curves with a constant FPF differ from the traditional contrast curves that fix the detection threshold as a function of separation and thereby allow the FPF to vary. Any image positions for which the specified FPF could not be reached due to self-subtraction were subsequently ignored. 
 
In the end, the contrast curves from the PCA and ADI PSF subtractions were combined and then converted to mass limit curves assuming the star's age and \Lp magnitude and using the COND evolutionary models \citep{2003A&A...402..701B}. The COND model was chosen primarily for convenience: many of the available models do not go down to low enough effective temperatures for the \Lp contrast limits reached in this work. We note that although the COND model assumes a hot start for the planet formation, the different predictions of the hot and cold start models have converged for the ages of the stars in the sample and for the mass limits we achieve \citep[see][]{2007ApJ...655..541M}.

\begin{table}
\caption{PCA PSF subtraction parameters\label{tab:psf_params} }
\centering
\begin{tabular}{lcccc}
  \hline \hline 
 Star & Stacking\tablefootmark{a} & \multicolumn{2}{c}{PCs used\tablefootmark{b}} & Mask size\tablefootmark{c}  \\ 
      &        & (\%) & (\#) &  (arcsec)  \\ 
 \hline 
$\tau$ Ceti & none & 10 & 414 &  0.30 \\ 
Kapteyn's star & none & 1 & 28 &  0.37 \\ 
AX mic & 10 & 2.5 & 9 &  0.40 \\ 
40 Eri & none & 0.5 & 6 &  0.27 \\ 
HD 36395 & 20 & 7.5 & 14 &  0.32 \\ 
HD 42581 & none & 5 & 186 &  0.10 \\ 
\hline 
\end{tabular} 
\tablefoot{ 
\tablefoottext{a}{Number of frames that were stacked prior to PSF modeling and subtraction, or none if no stacking was performed.} \\ 
\tablefoottext{b}{Quantity of principal components used to model the stellar PSF, expressed as a percentage of the total number frames or as the number of components used.} \\ 
\tablefoottext{c}{Radius of the central mask.} 
} 
\end{table}

\subsection{Radial velocity mass limits}
\label{sec:rv_lims}
To determine whether a planet with a given mass and semi-major axis would be detected by the RV data, we consider the overall scatter of the RV data and the epochs of the RV observations. For a given planet, we predict the stellar RV at each observation epoch and find the maximum difference in RV measured across the observations. If this maximum difference in the predicted RV at the observations epochs is larger than 5 times the standard deviation of the measured RVs (reported in Table \ref{tab:RV_obs}), then the planet signal is considered detectable by the RV data. 
We use this criterion for an RV detection instead of one derived from the typical Lomb-Scargle periodogram \citep{1976Ap&SS..39..447L,1982ApJ...263..835S} because the semi-major axes of interest correspond to orbital periods that are generally longer than the RV time baseline (for $a$ > 1 -- 6 AU), and thus these RV signals would not be periodic in the RV data. For the largest semi-major axes explored here of 50 -- 70 AU only 1 -- 5\% of the orbit is covered by the RV time baseline, making a periodogram a poor method for detecting these RV signals.

Prior to determining the standard deviation of the measured RVs, we consider the impact of known planets on the measured RV signal. Three of the six stars in the sample have had planets detected in their RV data with minimum masses in the super Earth/mini Neptune range: 
$\tau$ Ceti with 4 planets \citep[1.8 -- 3.9 $M_{\oplus}$,][]{2013A&A...551A..79T,2017AJ....154..135F}, 
Kapteyn's star with 2 planets (4.8 and 7 $M_{\oplus}$, \citealp{2014MNRAS.443L..89A}; but \citealp{2015ApJ...805L..22R} found that the lower mass planet signal is due to stellar activity) 
and 40 Eri with 1 planet \citep[8.5 $M_{\oplus}$,][]{2018MNRAS.480.2411M}.
All the measured RV semi-amplitudes for these planets are a factor of 3 -- 6 lower than the overall scatter measured in the data, so we do not expect these signals to change the detectabilty of the long orbital period companions as determined by the criterion above.  
\citet{2014MNRAS.441.1545T} also reported a planet with a mass of 1.9 times the mass of Neptune around HD 42581 based on combined UVES and HARPS RV data; however, we cannot reproduce the planet signal with the HARPS and Keck/HIRES RV data used in this work. We therefore analyze the HD 42581 data without removing a planetary signal, but we note that if this planet is real then the RV constraints for this target would improve. 

The brown dwarf companion to HD 42581 \citep[mass = 29 -- 39 $M_{\textrm{Jup}}$;][]{2015AJ....150...53N} could also have an effect on the RV signal of this star. This effect would most likley be in the form of a linear RV trend given the companion's wide separation and the time baseline of the RV data, with the slope of this trend highly depending on the companion's orbital parameters. Considering only circular orbits and given the approximate separation of HD 42581 B in the NACO data ($\sim$40 AU), the linear trend could be as high as $\sim$2 m/s/year; however, we do not find evidence for such a trend in the RV data. The non-detection of an RV signal is consistent with companion orbits that have semi-major axes larger than about 80 -- 100 AU when using the same detection criterion used above (i.e., the maximum difference in RV being larger than 5 times the overall RV standard deviation).

\subsection{Combining constraints from high contrast imaging and RV}

To combine the constraints on planetary architecture from HCI and RV, we employ a Monte Carlo approach to find the percentage of companions that would be detected for a given mass and semi-major axis ($a$). We explored a grid of companion masses ranging from 1 to 50 \MJ and semi-major axes from 1 to 45 AU (37 AU for 40 Eri, adjusted for the smaller FOV of this data). 
For each combination of mass and semi-major axis values, we randomly sample the other orbital parameters and then determine whether a planetary/substellar companion with this mass and orbit would have been detected with each method. Only circular orbits are considered ($e$ = 0), and the longitude of the ascending node ($\Omega$) is fixed because this orbital parameter does not affect the RV signal or the projected separation between the star and the companion in the HCI data. The time of closest approach ($T_0$) is also fixed to an arbitrary value, since this parameter is degenerate with the argument of periapse ($\omega$) for circular orbits. The mass of the star is set to the literature value given in Table~\ref{tab:stellar_prop}. For each combination of planet mass and semi-major axis, we sample the remaining orbital parameters of inclination ($i$; from a sin($i$) distribution for randomly oriented orbital planes) and $\omega$ (from a uniform distribution spanning 0 to 360 deg) 10,000 times.

For each of the 10,000 random orbits, the predicted projected separation at the epoch of the HCI observation and the predicted RV signal at the epochs of all RV measurements are found. 
To determine whether a companion on a given orbit would have been detected in the HCI data, we compare the companion mass to the mass limit curve at the projected separation. If the mass is higher than the mass limit, this planet is considered detectable by the HCI data. 
We apply the criteria outlined in Sec.~\ref{sec:rv_lims} to determine if the same planet is detectable by the RV data alone. 
Finally, we find the percentage of planets detectable by each method to measure the completeness for each combination of mass and semi-major axis.

\section{Results}
\label{sec:results}

\subsection{Contrast and mass limits from HCI data}
\label{sec:hcialone}

\begin{figure*}[t]
	\begin{center}
	\includegraphics[width=\textwidth]{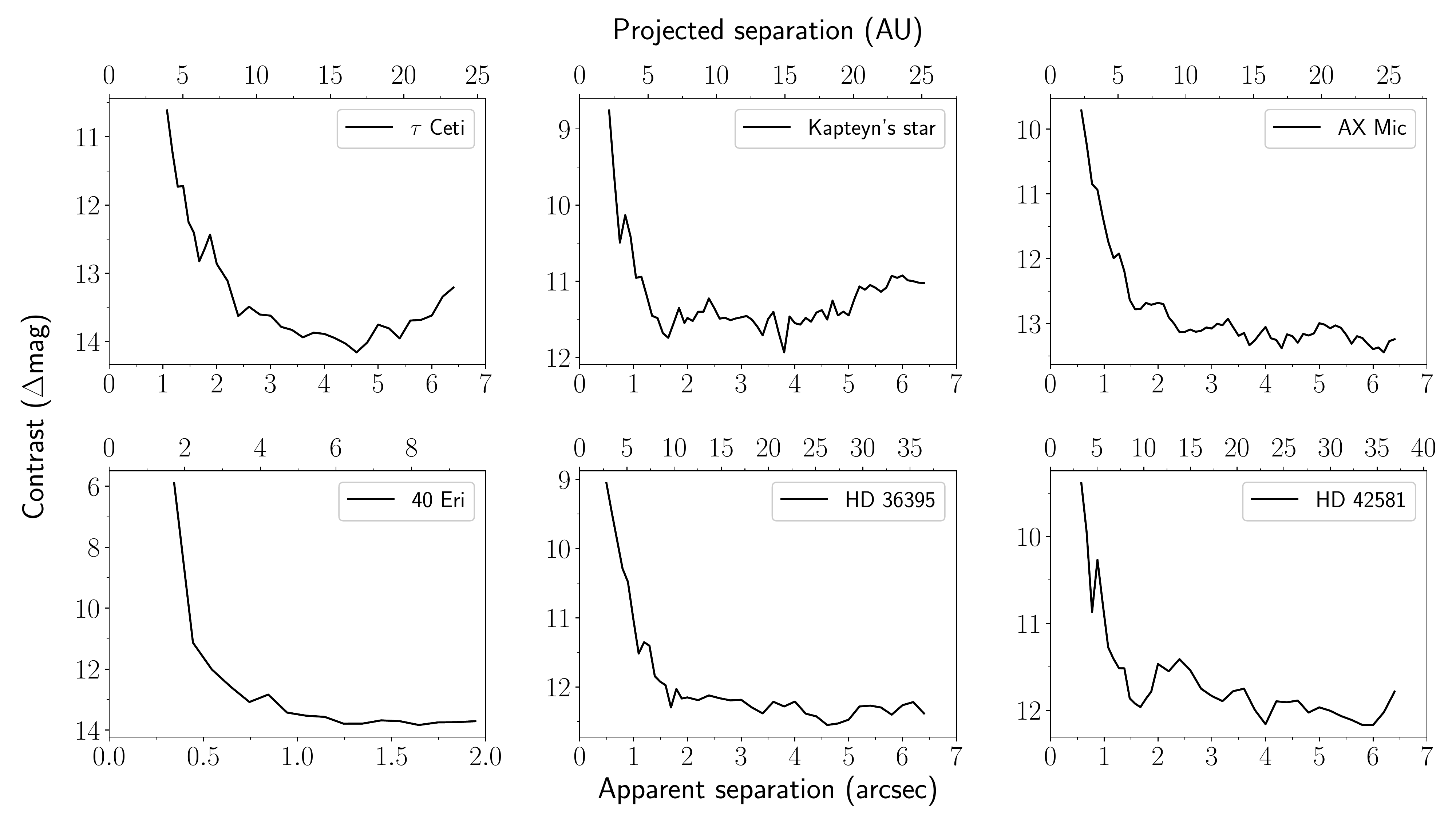}  
	\end{center}
	\caption
	{Limiting contrast curves for each star in the sample. The line plotted here is the contrast for a fixed FPF of $2.9 \times 10^{-7}$, which corresponds to a 5-sigma detection threshold for an underlying Gaussian distribution and for large sample sizes. These curves differ from the classical contrast curves since these give contrast for a fixed 5-sigma detection threshold and thus the FPF changes at close separations due to small number statistics \citep[see][]{2014ApJ...792...97M}. The limiting contrast shown here is an average of the contrast at 3 evenly spaced position angles (0, 120, and 240 degrees). The typical standard deviation of the contrasts computed at these 3 position angles is 0.15 mag for a given separation. We do not plot limiting contrasts for the separations at which the specified FPF could not be achieved due to self-subtraction. 
	\label{fig:contrast_curves}
    }
\end{figure*}

\begin{figure*}[t]
	\begin{center}
	\includegraphics[width=\textwidth]{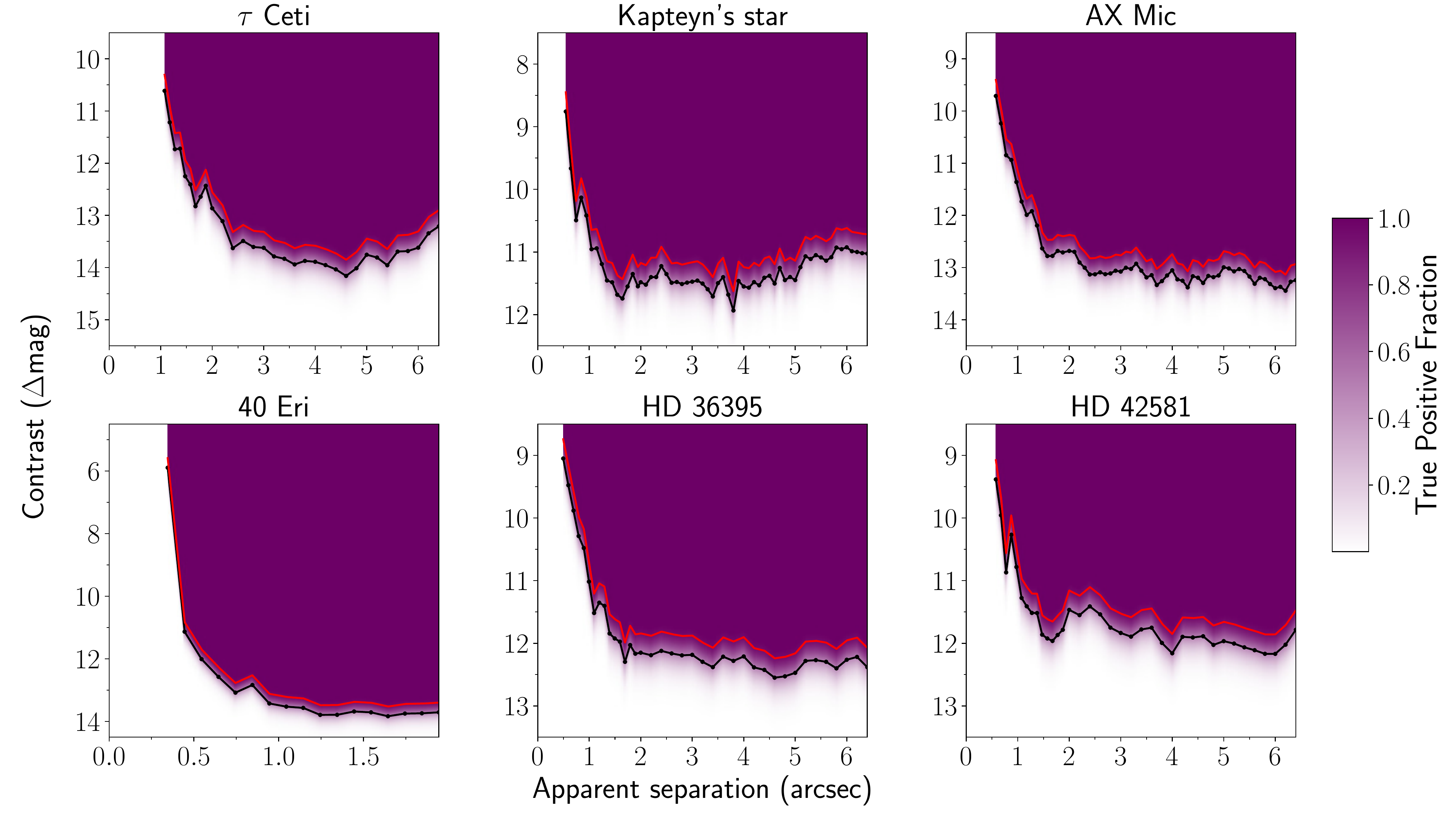}
	\end{center}
	\caption
	{ Performance maps for all stars in the sample, after \citet{2018AJ....155...19J}. The colors show the TPF as a function of contrast and projected separation for a fixed FPF of $2.9 \times 10^{-7}$ (5-sigma threshold for Gaussian statistics). The red line is the 95\% completeness curve and the black line is the 50\% completeness curve (as plotted in Fig.\ref{fig:contrast_curves}).
	\label{fig:performance_maps}
    }
\end{figure*}

The limiting \Lp contrast curves for all stars in the sample are shown in Fig.~\ref{fig:contrast_curves}. We constrain the presence of companions out to projected separations of 2.0\arcsec~for 40 Eri and to 6.5\arcsec~for the remaining targets. These angular separations correspond to physical separations of 10 AU and 25 -- 35 AU, respectively, for the very nearby distances of these stars. The limiting contrasts for a fixed FPF of $2.9 \times 10^{-7}$ go down to $\sim$14 magnitudes in the background-limited regions, which are at angular separations of 1 -- 3\arcsec~depending on the brightness of the target star. 
To explore the completeness of the NACO HCI data alone, we also present performance maps in Fig.~\ref{fig:performance_maps}, following the method described in \citet{2018AJ....155...19J}. These performance maps present the true positive fraction (TPF) as a function of separation and contrast for the fixed FPF of $2.9 \times 10^{-7}$, allowing a more thorough exploration of the trade-off between FPF and TPF. The 50\% completeness curve is highlighted, which is the same completeness shown in Fig.~\ref{fig:contrast_curves} by definition, as well as the 95\% completeness. In both the final contrast curves and performance maps, separations for which the specified FPF could not be reached due to self-subtraction are not plotted. 

\begin{figure*}[t]
	\begin{center}
	\includegraphics[width=\textwidth]{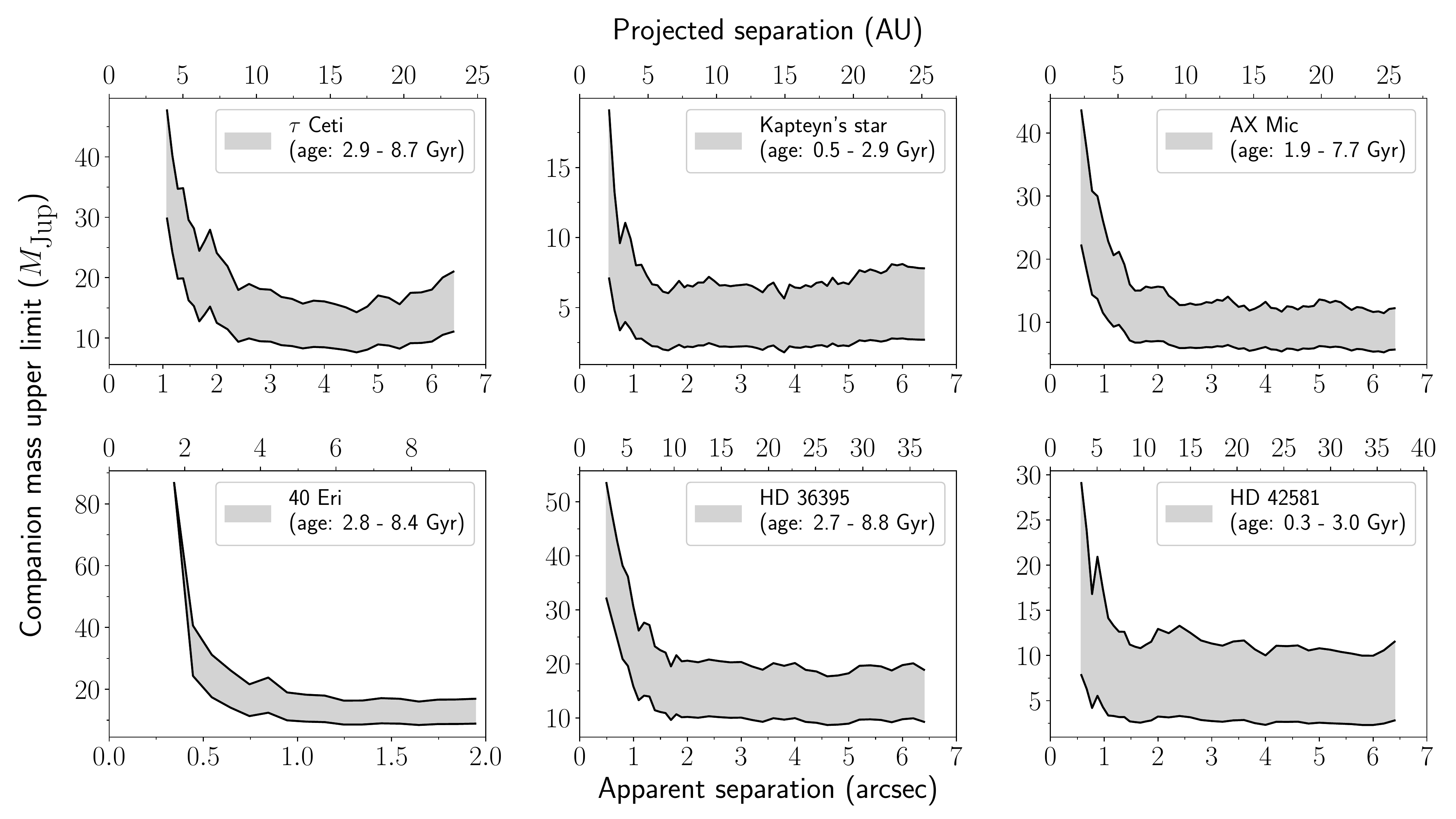}  
	\end{center}
	\caption
	{ Limiting companion mass curves for all stars in the sample. The spread in the limiting masses is due to the uncertainty in the age of the star. In the background limit, where direct imaging adds the most information compared to RV. The limits reach down to a few \MJ for the stars with the youngest ages and $\sim$20 \MJ for the oldest ages.
	\label{fig:mass_lims}
    }
\end{figure*}

Fig.~\ref{fig:mass_lims} shows the 50\% completeness contrast limits converted into mass limits given the star's age, distance, and apparent magnitude. The ages of the stars are quite uncertain with typical errors of $\sim$50 -- 70\%, and this results in a spread in the limiting mass of about 10 \MJ. The mass limits reach down to 2 -- 10 \MJ for the stars with youngest ages (<3 Gyr; Kapteyn's star and HD 42581). The remaining stars are likely to be older than 2 -- 3 Gyr, resulting in mass limits of 5 -- 10 \MJ in the background limit, assuming the youngest possible ages, or 15 -- 20 \MJ, assuming the oldest possible ages up to 9 Gyr. 

\subsection{Planetary architecture constraints from combining HCI and RV}
\label{sec:hciandrv}

\begin{figure*}[t]
	\begin{center}
	\includegraphics[width=\textwidth]{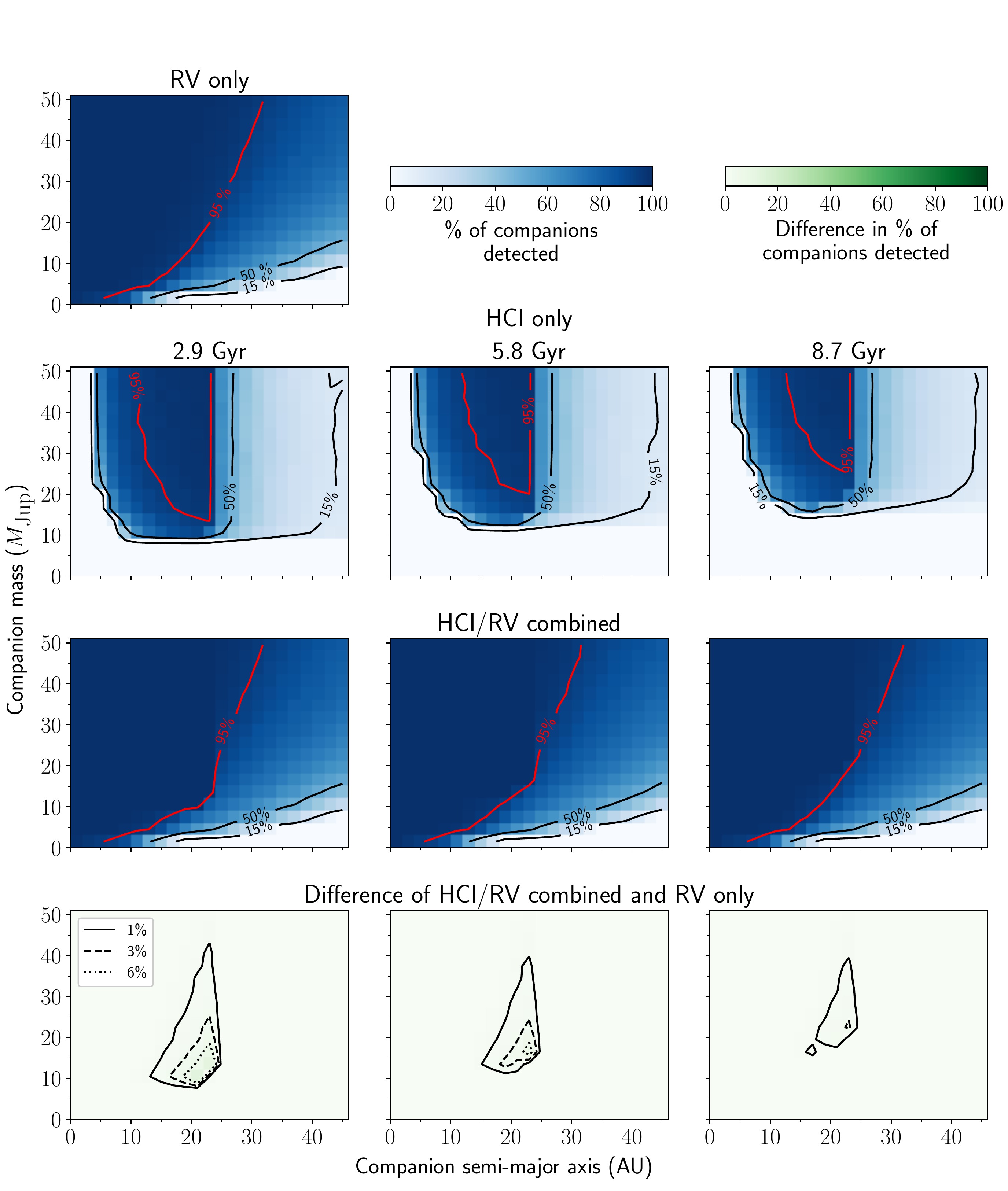}
	\end{center}
	\caption
	{ Constraints on companion mass and semi-major axis for $\tau$ Ceti. The percentage of companions that can be detected by RV alone is plotted on the top row, the percentage of companions that can be detected by HCI alone is plotted in the second row (assuming the minimum, nominal, and maximum ages of the star), the combined HCI/RV constraints in the third row, and the difference between the HCI/RV combined and the RV only constraints in the fourth row. 
	Semi-major axes corresponding to 2 times the projected physical separation probed by the NACO field of view are shown. For $\tau$ Ceti, we find that including the HCI constraints results in only up to 13\% more detectable companions for masses of 10 -- 40 \MJ and semi-major axes of 15 -- 25 AU. 
	\label{fig:tau_ceti_lims}
    }
\end{figure*}

\begin{figure*}[t]
	\begin{center}
	\includegraphics[width=\textwidth]{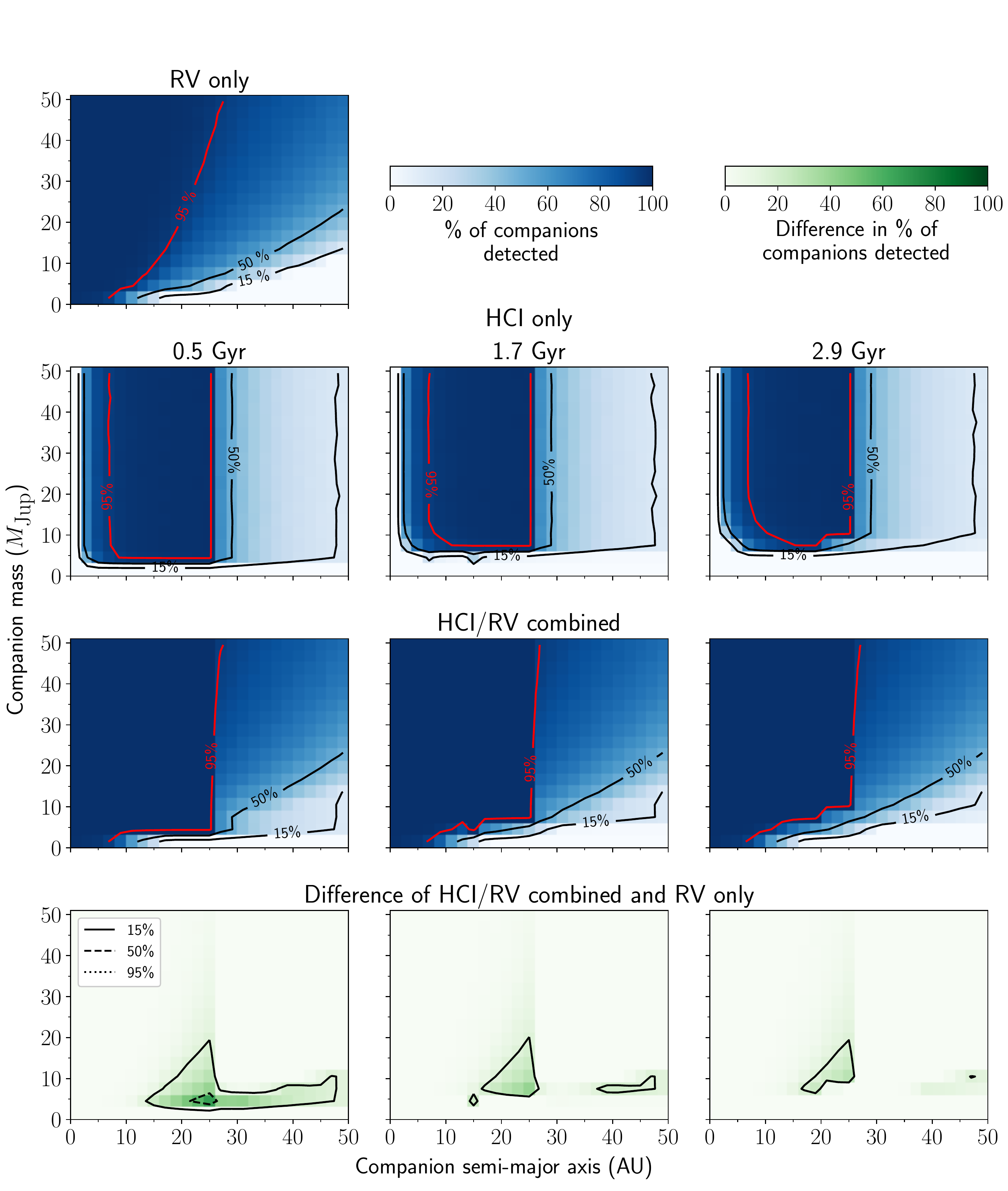}
	\end{center}
	\caption
	{ Constraints on companion mass and semi-major axis for Kapteyn's star. HCI adds significant information (up to 68\% more companions that can be detected for the minimum age and up to 29\% for the maximum age), concentrated at companion masses of 5 -- 10 \MJ and semi-major axes of 15 -- 25 AU. The detectable companions also increase by up to 41\% for the minimum age and 16\% for the maximum age at semi-major axes larger than the NACO field of view.
	\label{fig:kapteyns_lims}
    }
\end{figure*}

\begin{figure*}[t]
	\begin{center}
	\includegraphics[width=\textwidth]{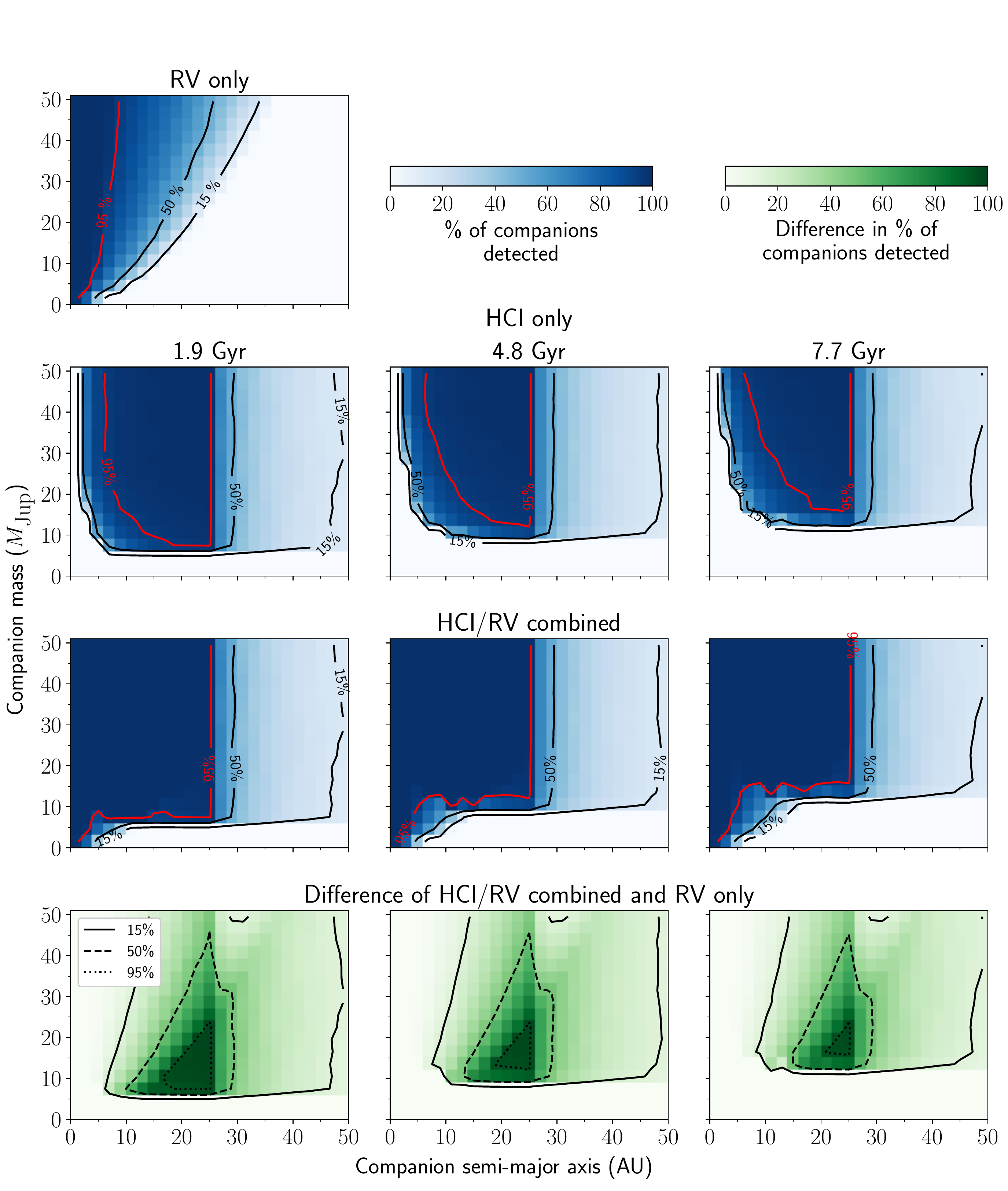}
	\end{center}
	\caption
	{ Constraints on companion mass and semi-major axis for AX Mic. HCI adds significant information (up to 99\% more companions that can be detected even for the maximum age), concentrated at companion masses of 8 -- 14 \MJ and semi-major axes of 15 -- 25 AU. The detectable companions also increase by up to 65\% at semi-major axes larger than the NACO field of view for all ages. The RV data on this star are the least constraining in the sample due to the short times baseline (742 days) and small number of measurements (8 nights).
	\label{fig:axmic_lims}
    }
\end{figure*}

\begin{figure*}[t]
	\begin{center}
	\includegraphics[width=\textwidth]{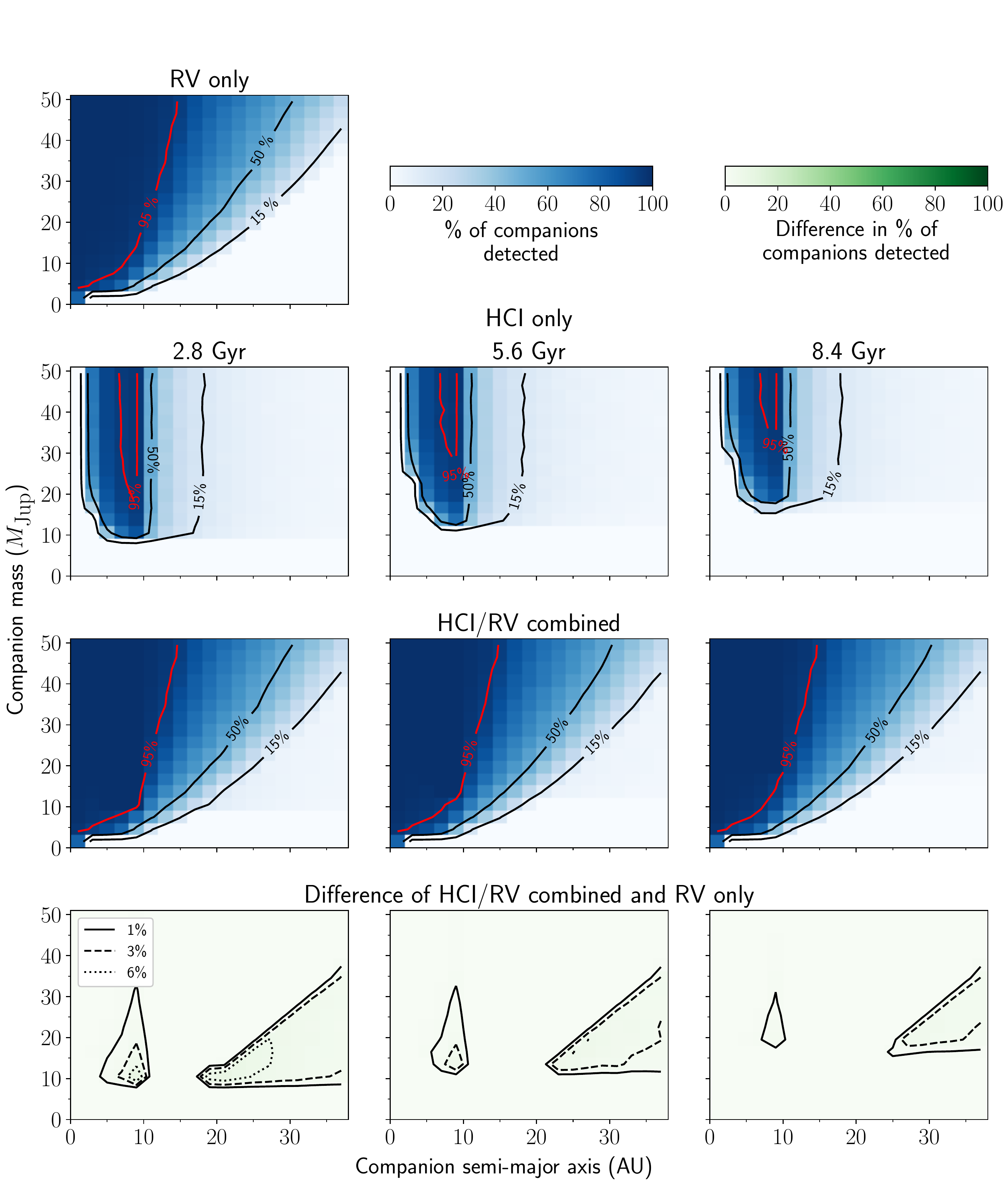}
	\end{center}
	\caption
	{ Constraints on companion mass and semi-major axis for 40 Eri. The HCI data only extend to projected physical separations of $\sim$10 AU, and for semi-major axes less than this value the percentage of companions that can be detected is increased by up to 9\% for the minimum age and 3\% for the maximum age. Semi-major axes up to 4 times the HCI field of view are shown, and we find that the HCI data increase the percentage of companions that can be detected by similar amounts for large semi-major axes $>$20 -- 25 AU. 
	\label{fig:40_eri_lims}
    }
\end{figure*}

\begin{figure*}[t]
	\begin{center}
	\includegraphics[width=\textwidth]{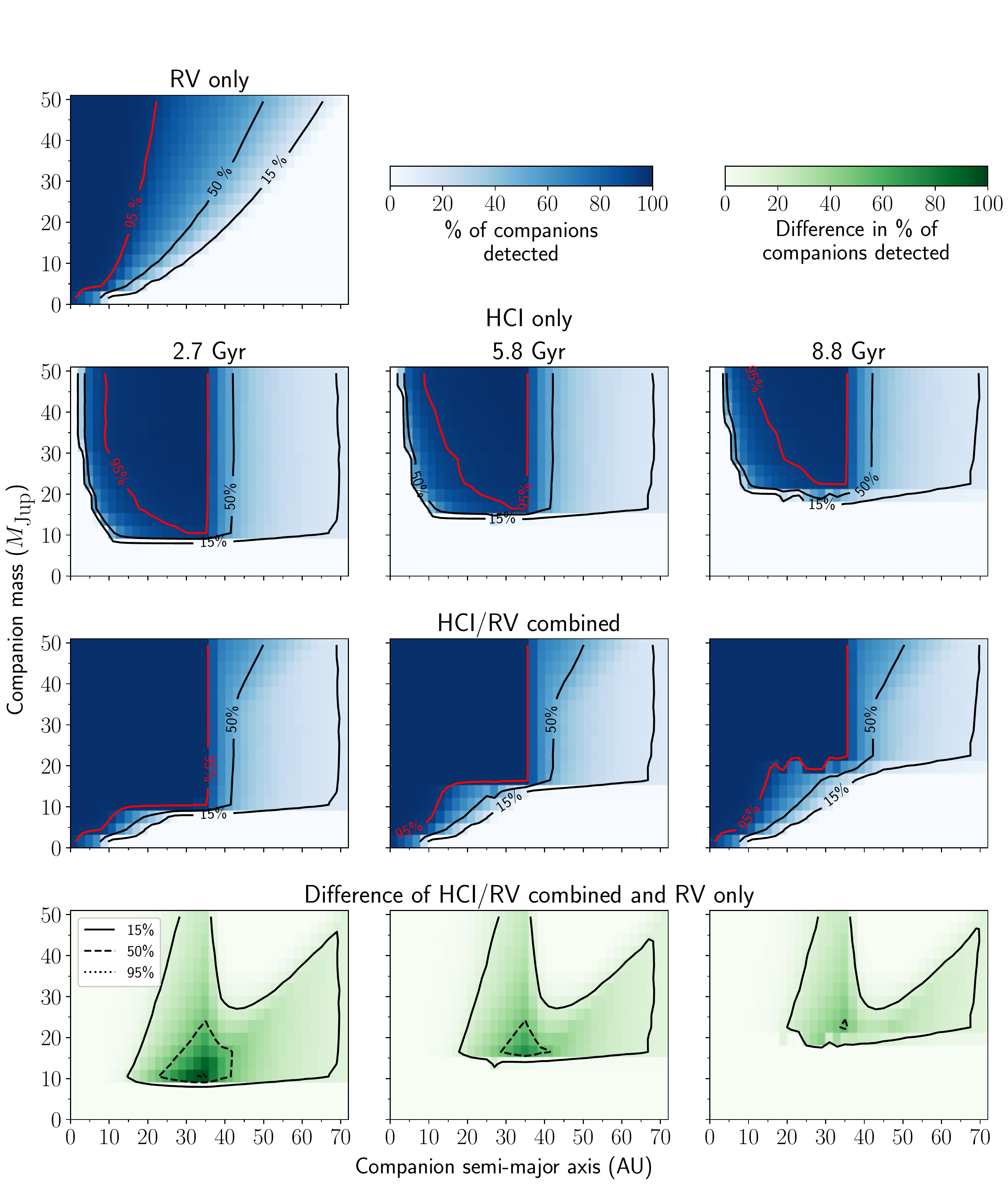}
	\end{center}
	\caption
	{ Constraints on companion mass and semi-major axis for HD 36395. HCI adds significant information (up to 95\% more companions that can be detected for the minimum age and up to 55\% for the maximum age) for companions masses of 10 -- 20 \MJ and semi-major axes of 25 -- 35 AU. Information is also added at wider semi-major axes, increasing the percentage of companions that can be detected by up to 40 -- 80\% for semi-major axes just larger than the NACO FOV and by 15\% for semi-major axes of 60 -- 70 AU and for masses of 20 -- 50 \MJ.
	\label{fig:hd36395_lims}
    }
\end{figure*}

\begin{figure*}[t]
	\begin{center}
	\includegraphics[width=\textwidth]{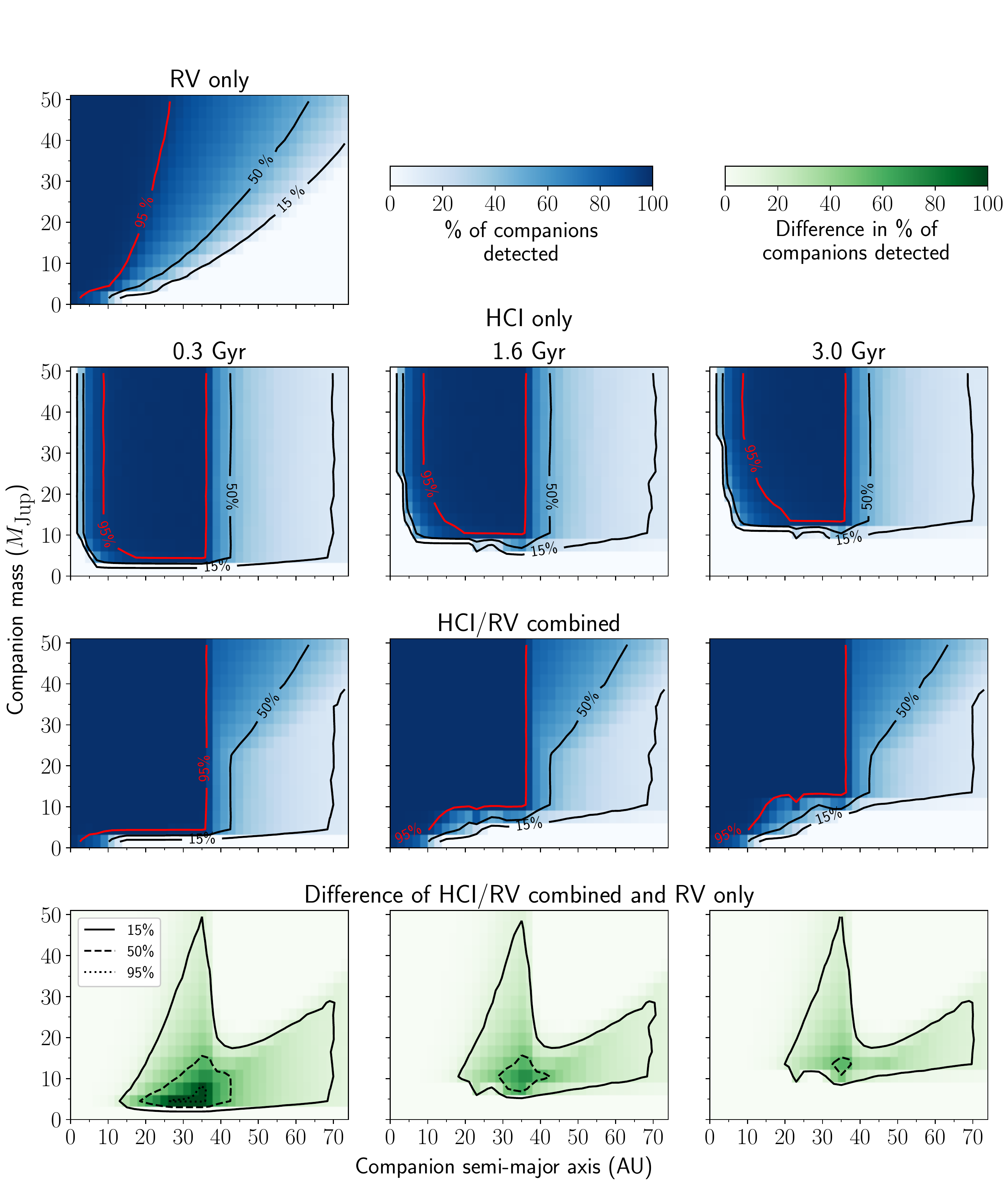}
	\end{center}
	\caption
	{ Constraints on companion mass and semi-major axis for HD 42581. HCI increases the companions that can be detected by up to \
	100\% for the minimum age and up to 56\% for the maximum age for masses of 5 -- 10 \MJ and semi-major axes of 27 -- 37 AU. Information is also added for wider separations up to 70 AU for masses from 5 -- 10 \MJ to 30 \MJ.
	\label{fig:hd42581_lims}
    }
\end{figure*}

The percentage of the 10,000 orbits for each mass/semi-major axis combination that can be detected by HCI, RV, and by either HCI or RV are plotted in Figs.~\ref{fig:tau_ceti_lims} -- \ref{fig:hd42581_lims}. Semi-major axes out to 2 times (4 times for 40 Eri) the maximum projected separation probed by the NACO FOV are explored, since orbits with $a$ greater than the FOV may also fall into the NACO frame depending on the orbital phase at the time of the observation. We focus on large semi-major axes that correspond to longer orbital periods than are fully covered by the RV data, since this is the parameter space where HCI could improve constraints when combined with RV. 
Over these semi-major axes, we find that the limiting mass of the RV data increases quickly as a function of semi-major axis, whereas the HCI constraints are relatively flat for semi-major axes corresponding to the projected physical separations of the background limit. 
For 4 out of 6 of the stars, HCI adds significant information to the RV constraints (AX mic, Kapteyn's star, HD 36395, HD 42581). 
The additional constraints from HCI are concentrated at low companion masses near the HCI detection limit (5 -- 20 \MJ) and large semi-major axes out to the maximum separation probed by the NACO FOV ($\sim$25 -- 40 AU). 
When including the information from HCI, the percentage of detectable substellar companions increases by up to 50 -- 90\% for Kapteyn's star, HD 36395, and HD 42581 depending on the age assumed. For AX mic, which is the star with the fewest RV measurements (8 measurements after nightly binning) and the shortest RV time baseline (741 days), the percentage of companions that can be detected increases by up to 99\% irrespective of the stellar age.

For the stars $\tau$ Ceti and 40 Eri, the addition of HCI data increases the detectable orbits by at most 10\%. The HCI mass limits for 40 Eri only go out to projected physical separations of $\sim$10 AU, and the RV data provide strong constraints for semi-major axes smaller than this value despite the relatively large scatter in the data (12.4 m/s). 
$\tau$ Ceti is well known as an extremely stable RV star, and of the stars in this sample it has the longest RV time baseline (7,354 days or $\sim$19 years) and lowest RV scatter (2.8 m/s). Thus the RV data of this star place very powerful constraints on the planetary architecture even at wide semi-major axes up to 40 AU (orbital periods of over 300 years). 
Still, for 40 Eri and $\tau$ Ceti, HCI can detect a few tens of the 10,000 sampled orbits that RV misses for small companion masses and large semi-major axes. 
The orbits at which HCI can detect planets tend to have small inclinations (less than 20 -- 30 deg) and projected separations close to the maximum possible value at the time of the HCI observations. 
It is possible to detect these companions in HCI but not in RV because at these large semi-major axes, the RV time baseline only covers a small fraction of the orbit (down to 13\% for $a$ < the NACO FOV) and thus the smallest changes of RV over the baseline occur around the times of maximum and minimum RV value, where the derivative of the RV is 0.
This is also the section of the orbit where the projected separation is the highest, allowing some companions to be detectable by HCI.  

HCI can even add information for semi-major axes beyond the extent of the image for lower inclinations, due to the fact that there is a fraction of orbits for which the projected separation at the time of the HCI observation still falls in the FOV. These constraints are as high as 40 -- 80\% for the semi-major axes just larger than the FOV and steadily drop with increasing semi-major axis. The triangular shape of the additional HCI constraints is set by the mass limit in the HCI background-limited regime at the lower end and by the RV mass limit at the upper end. 
The power of HCI to add information for semi-major axes beyond the observed projected separations is amplified by the fact that higher inclinations are more likely given randomly-oriented orbital planes. For the star 40 Eri, HCI can even detect companions with semi-major axes up to 4 times the extent of the NACO FOV, depending on the planet's orbital phase at the time of the observations, despite the smaller projected separations probed in the imaging data of this target.

\section{Discussion}
\label{sec:disc}

We have found that by combining constraints from NACO \Lp HCI with long-term RV monitoring data, the constraints on the planetary architecture of nearby stars are greatly improved, primarily for substellar masses of 5 -- 20 \MJ and for semi-major axes similar to those of the ice giants in our own solar system (15 -- 40 AU).  The amount of improvement from including the HCI data depends primarily on the age of the star, with ages <5 Gyr leading to mass limits in the planetary mass regime. 
Differences in the RV data time baseline and scatter can also significantly change the constraints from these data and thus how much information HCI can add. 
For example, the difference between the contributions from HCI for HD 36395 and HD 42581 for the age of $\sim$3 Gyr are driven by the longer time baseline (7,240 days versus 6,220 days) and the lower RV scatter (4.4 versus 6.0 m/s) of HD 42581, resulting in HCI detecting up to 96\% more companions for HD 36395 compared to up to 56\% for HD 42581.

The fact that we did not detect any additional companions in the archival HCI is perhaps not surprising given the current statistics on giant planets and substellar companions at wide separations. HCI surveys have found that giant planets on wide orbits are relatively rare: planets of 5 -- 13 \MJ and $a$ = 30 -- 300 AU are found around less than $\sim$4\% of FGK and M stars \citep{2016PASP..128j2001B}. Combined statistics on M stars from RV, HCI, and microlensing surveys yield similarly low giant planet occurrence rates of 1.5 -- 3\% for planets of 1 -- 10 \MJ and $a$ = 10 -- 1000 AU \citep{2018A&A...612L...3M}. However, we note that the semi-major axes probed for the nearby stars in this sample are smaller than those probed by the typical HCI survey, since those surveys focus on young stars that are generally found in moving groups at around 30 -- 50 pc compared to the <6 pc distances explored in this work. The intermediate semi-major axis range of 15 -- 40 AU probed by HCI here is also beyond those semi-major axes well constrained by RV surveys, which give statistical constraints out to a few AU \citep[e.g.,][]{2010PASP..122..905J,2013A&A...549A.109B}. 
While this sample of 6 stars does not allow us to draw any significant conclusions about the statistics of substellar companions at these intermediate semi-major axes, it does highlight the new information gained by direct imaging searches around nearby stars both now and with future facilities, despite the generally older ages of field stars.

The HCI and RV limits reported here assume that all possible orbital planes are equally likely and thus use no prior information on the inclination of the possible planetary system. However $\tau$ Ceti, for example, hosts a debris disk that was found to have an inclination of 35 $\pm$ 10 deg by \textit{Herschel Space Observatory} imaging \citep{2014MNRAS.444.2665L}. Assuming this inclination for any possible planetary companions would reduce the strength of their RV signal significantly (57\% of the RV signal from the same planet with an edge-on orbit) and would also increase the likelihood that a companion would fall at a wide separation in the single epoch of HCI data. This shows that for future HCI observations of other nearby stars, prior information on the inclination of the system such as debris disk inclination or stellar rotation axis could be taken into account to improve the information gained by adding HCI to existing long-term RV monitoring \citep{2010MNRAS.408..514J}. 

In addition to being favorable targets for HCI and RV observations, nearby stars such as those in this sample will also be prime targets for astrometric searches for planets with \textit{Gaia} \citep[e.g.,][]{2014ApJ...797...14P}. 
With an astrometric precision on the order of tens of $\mu$as for V band magnitudes down to 12\footnote{\url{https://www.cosmos.esa.int/web/gaia/science-performance} gives the nominal astrometric performance of \textit{Gaia} (site accessed on 17 April 2019).}, \textit{Gaia} will be especially sensitive to planets on wide orbits around the nearest stars for orbital periods well covered by the nominal \textit{Gaia} mission time baseline of 5 years. 
\textit{Gaia} will be less sensitive to planets on orbital periods longer than the mission time baseline, since it is then more challenging to disentangle the planet signal from the parallax and proper motion signals \citep[see][]{2018A&A...614A..30R}. 
Therefore astrometry with \textit{Gaia} will probe the parameter space of a < 5 -- 6 AU for masses down to a few Neptune masses for the very nearby stars in this sample. However, further analysis will be required to determine what planet masses will detectable for much longer orbital periods where HCI currently adds the most information. 
The astrometry for very bright stars (G < 6 mag, $\tau$ Ceti and 40 Eri in this sample) in the \textit{Gaia} DR2 is also currently limited by calibration issues from CCD saturation effects; however, the calibration is expected to improve in future data releases \citep{2018A&A...616A...2L}. 

Recent works have also combined proper motion and/or position information from the \textit{Hipparcos} and \textit{Gaia} catalogues, leveraging the $\sim$24 year time baseline between these two missions to constrain the masses of known companions \citep{2018A&A...615A.149C,2018NatAs...2..883S, 2018arXiv181107285B,2019ApJ...871L...4D} or search for new companions \citep{2018arXiv181108902K}. The latter work specifically targeted nearby stars (< 50 pc) to search for "proper motion anomalies": the difference between the long-term proper motion obtained from the positions in the \textit{Hipparcos} and \textit{Gaia} catalogues and the proper motion from either catalogue due to the presence of a companion. 
That work presents proper motion anomaly measurements for all stars in this sample except HD 36395, and found a significant (>5$\sigma$) change in proper motion only for HD 42581, which is due to its brown dwarf companion HD 42581 B. 
We note that the error bars on the measured proper motion anomalies may be underestimated in that work given the results of \citet{2018ApJS..239...31B}, which found that combining the Hippcarcos and \textit{Gaia} catalogues in the same reference frame requires inflating the proper motion and position errors. 

The upcoming JWST will be a powerful tool for HCI exoplanetary science thanks to the coronagraphs available in the NIRCam and MIRI instruments \citep{2005AdSpR..36.1099B,2010PASP..122..162B}.
We explored the mass limits that can be reached with JWST/NIRCam for the stars in this sample, specifically for the separations at which the sensitive background limit should be reached (approximately 3 -- 5\arcsec~or 10 -- 30 AU). 
Based on the evolutionary and atmospheric models presented in \citet{2019A&A...623A..85L} and assuming an effective integration time of $\sim$30 min, we find that planets with masses down to 1 -- 2 \MJ would be detected by JWST for these nearby field stars, roughly an order of magnitude better than the mass limits presented here. With these mass limits, the JWST HCI data could detect Jupiter-mass planets currently hidden in the RV data, even for stars with very long time baselines and small overall RV scatter like $\tau$ Ceti. 

In the near future, new HCI observations of these targets from the ground could also improve the imaging constraints presented in this work. The NACO data analyzed here have relatively short integration times of $\sim$10 -- 25 min, and increasing the total integration time would improve the mass limits at wide separations by a few \MJ.  The upcoming ERIS instrument for the VLT will provide improved 1 -- 5 $\mu$m HCI capabilities compared to NACO when it sees first light in 2020 \citep{2018SPIE10702E..09D,2018SPIE10702E..46K}. ERIS will use the AO Facility of the UT4 telescope with its deformable secondary mirror, resulting in fewer warm reflections before the light enters the cryogenic camera compared to NACO and improved sensitivity in the background limit. Finally, having a $\sim$12 years time baseline between these NACO observations and any new HCI observations would improve completeness due to the orbital motion of planets between the two epochs. Planets on certain orbits will have moved significantly on the sky, such that a planet initially too close to the host star would be further away and more easily detected in the later epoch. This effect is stronger for nearby stars like those in this sample because the HCI data can probe smaller semi-major axes and shorter orbital periods compared to more distant systems.

\section{Summary}

We quantified the limits on possible planetary architectures around 6 very nearby stars (d < 6 pc) by combining limits on the mass and semi-major axes of possible companions from HCI and from RV measurements. We find that HCI adds information to the long-term RV monitoring constraints for intermediate semi-major axes (15 -- 40 AU) and masses down to the background limit of the archival \Lp NACO imaging data reanalyzed here (5 -- 20 \MJ). 
The completeness was improved significantly for 4 of the 6 stars in the sample, and even by up to a few percent for $\tau$ Ceti, the star with the longest RV time baseline and lowest scatter, specifically for companions on nearly face-on orbits ($i$ < 30 deg) and for cases where the RV time window included the time when the derivative of the RV signal is 0. 
These results show the power of combining constraints from different and complementary exoplanet detection methods. In the future, combining information from these the HCI and RV methods as well as from \textit{Gaia} astrometry and JWST HCI observations will allow us to fully characterize our knowledge of the substellar companion population in the solar neighborhood at intermediate semi-major axes. The resulting information can be used to inform the target lists for future space missions like HabEx and LUVOIR, which will look for Earth analogues around the nearest stars. 

\begin{acknowledgements}
This work has been carried out within the framework of the National Center for Competence in Research PlanetS supported by the SNSF. 
\end{acknowledgements}

\bibliographystyle{aa}
\bibliography{exoplanets}

\begin{thebibliography}{78}
\expandafter\ifx\csname natexlab\endcsname\relax\def\natexlab#1{#1}\fi

\bibitem[{{Amara} \& {Quanz}(2012)}]{2012MNRAS.427..948A}
{Amara}, A. \& {Quanz}, S.~P. 2012, \mnras, 427, 948

\bibitem[{{Anglada-Escude} {et~al.}(2014){Anglada-Escude}, {Arriagada},
  {Tuomi}, {Zechmeister}, {Jenkins}, {Ofir}, {Dreizler}, {Gerlach}, {Marvin},
  {Reiners}, {Jeffers}, {Butler}, {Vogt}, {Amado}, {Rodriguez-Lopez},
  {Berdinas}, {Morin}, {Crane}, {Shectman}, {Thompson}, {Diaz}, {Rivera},
  {Sarmiento}, \& {Jones}}]{2014MNRAS.443L..89A}
{Anglada-Escude}, G., {Arriagada}, P., {Tuomi}, M., {et~al.} 2014, \mnras, 443,
  L89

\bibitem[{{Astudillo-Defru} {et~al.}(2017){Astudillo-Defru}, {Delfosse},
  {Bonfils}, {Forveille}, {Lovis}, \& {Rameau}}]{2017A&A...600A..13A}
{Astudillo-Defru}, N., {Delfosse}, X., {Bonfils}, X., {et~al.} 2017, \aap, 600,
  A13

\bibitem[{{Baraffe} {et~al.}(2003){Baraffe}, {Chabrier}, {Barman}, {Allard}, \&
  {Hauschildt}}]{2003A&A...402..701B}
{Baraffe}, I., {Chabrier}, G., {Barman}, T.~S., {Allard}, F., \& {Hauschildt},
  P.~H. 2003, \aap, 402, 701

\bibitem[{{Beichman} {et~al.}(2010){Beichman}, {Krist}, {Trauger}, {Greene},
  {Oppenheimer}, {Sivaramakrishnan}, {Doyon}, {Boccaletti}, {Barman}, \&
  {Rieke}}]{2010PASP..122..162B}
{Beichman}, C.~A., {Krist}, J., {Trauger}, J.~T., {et~al.} 2010, \pasp, 122,
  162

\bibitem[{{Beuzit} {et~al.}(2019){Beuzit}, {Vigan}, {Mouillet}, {Dohlen},
  {Gratton}, {Boccaletti}, {Sauvage}, {Schmid}, {Langlois}, {Petit},
  {Baruffolo}, {Feldt}, {Milli}, {Wahhaj}, {Abe}, {Anselmi}, {Antichi},
  {Barette}, {Baudrand}, {Baudoz}, {Bazzon}, {Bernardi}, {Blanchard}, {Brast},
  {Bruno}, {Buey}, {Carbillet}, {Carle}, {Cascone}, {Chapron}, {Chauvin},
  {Charton}, {Claudi}, {Costille}, {De Caprio}, {Delboulb{\'e}}, {Desidera},
  {Dominik}, {Downing}, {Dupuis}, {Fabron}, {Fantinel}, {Farisato},
  {Feautrier}, {Fedrigo}, {Fusco}, {Gigan}, {Ginski}, {Girard}, {Giro},
  {Gisler}, {Gluck}, {Gry}, {Henning}, {Hubin}, {Hugot}, {Incorvaia}, {Jaquet},
  {Kasper}, {Lagadec}, {Lagrange}, {Le Coroller}, {Le Mignant}, {Le Ruyet},
  {Lessio}, {Lizon}, {Llored}, {Lundin}, {Madec}, {Magnard}, {Marteaud},
  {Martinez}, {Maurel}, {M{\'e}nard}, {Mesa}, {M{\"o}ller-Nilsson}, {Moulin},
  {Moutou}, {Orign{\'e}}, {Parisot}, {Pavlov}, {Perret}, {Pragt}, {Puget},
  {Rabou}, {Ramos}, {Reess}, {Rigal}, {Rochat}, {Roelfsema}, {Rousset}, {Roux},
  {Saisse}, {Salasnich}, {Santambrogio}, {Scuderi}, {Segransan}, {Sevin},
  {Siebenmorgen}, {Soenke}, {Stadler}, {Suarez}, {Tiph{\`e}ne}, {Turatto},
  {Udry}, {Vakili}, {Waters}, {Weber}, {Wildi}, {Zins}, \&
  {Zurlo}}]{2019arXiv190204080B}
{Beuzit}, J.~L., {Vigan}, A., {Mouillet}, D., {et~al.} 2019, arXiv e-prints,
  arXiv:1902.04080

\bibitem[{{Boccaletti} {et~al.}(2005){Boccaletti}, {Baudoz}, {Baudrand},
  {Reess}, \& {Rouan}}]{2005AdSpR..36.1099B}
{Boccaletti}, A., {Baudoz}, P., {Baudrand}, J., {Reess}, J.~M., \& {Rouan}, D.
  2005, Advances in Space Research, 36, 1099

\bibitem[{{Bonfils} {et~al.}(2013){Bonfils}, {Delfosse}, {Udry}, {Forveille},
  {Mayor}, {Perrier}, {Bouchy}, {Gillon}, {Lovis}, {Pepe}, {Queloz}, {Santos},
  {S{\'e}gransan}, \& {Bertaux}}]{2013A&A...549A.109B}
{Bonfils}, X., {Delfosse}, X., {Udry}, S., {et~al.} 2013, \aap, 549, A109

\bibitem[{{Bowler}(2016)}]{2016PASP..128j2001B}
{Bowler}, B.~P. 2016, Publications of the Astronomical Society of the Pacific,
  128, 102001

\bibitem[{{Bowler} {et~al.}(2018){Bowler}, {Dupuy}, {Endl}, {Cochran},
  {MacQueen}, {Fulton}, {Petigura}, {Howard}, {Hirsch}, {Kratter}, {Crepp},
  {Biller}, {Johnson}, \& {Wittenmyer}}]{2018AJ....155..159B}
{Bowler}, B.~P., {Dupuy}, T.~J., {Endl}, M., {et~al.} 2018, \aj, 155, 159

\bibitem[{{Brandl} {et~al.}(2018){Brandl}, {Absil}, {Ag{\'o}cs}, {Baccichet},
  {Bertram}, {Bettonvil}, {van Boekel}, {Burtscher}, {van Dishoeck}, {Feldt},
  {Garcia}, {Glasse}, {Glauser}, {G{\"u}del}, {Haupt}, {Kenworthy}, {Labadie},
  {Laun}, {Lesman}, {Pantin}, {Quanz}, {Snellen}, {Siebenmorgen}, \& {van
  Winckel}}]{2018SPIE10702E..1UB}
{Brandl}, B.~R., {Absil}, O., {Ag{\'o}cs}, T., {et~al.} 2018, in Society of
  Photo-Optical Instrumentation Engineers (SPIE) Conference Series, Vol. 10702,
  107021U

\bibitem[{{Brandt}(2018)}]{2018ApJS..239...31B}
{Brandt}, T.~D. 2018, The Astrophysical Journal Supplement Series, 239, 31

\bibitem[{{Brandt} {et~al.}(2018){Brandt}, {Dupuy}, \&
  {Bowler}}]{2018arXiv181107285B}
{Brandt}, T.~D., {Dupuy}, T., \& {Bowler}, B.~P. 2018, arXiv e-prints,
  arXiv:1811.07285

\bibitem[{{Butler} {et~al.}(2017){Butler}, {Vogt}, {Laughlin}, {Burt},
  {Rivera}, {Tuomi}, {Teske}, {Arriagada}, {Diaz}, {Holden}, \&
  {Keiser}}]{2017AJ....153..208B}
{Butler}, R.~P., {Vogt}, S.~S., {Laughlin}, G., {et~al.} 2017, \aj, 153, 208

\bibitem[{{Calissendorff} \& {Janson}(2018)}]{2018A&A...615A.149C}
{Calissendorff}, P. \& {Janson}, M. 2018, \aap, 615, A149

\bibitem[{{Cheetham} {et~al.}(2018){Cheetham}, {S{\'e}gransan}, {Peretti},
  {Delisle}, {Hagelberg}, {Beuzit}, {Forveille}, {Marmier}, {Udry}, \&
  {Wildi}}]{2018A&A...614A..16C}
{Cheetham}, A., {S{\'e}gransan}, D., {Peretti}, S., {et~al.} 2018, \aap, 614,
  A16

\bibitem[{{Crepp} {et~al.}(2012){Crepp}, {Johnson}, {Fischer}, {Howard},
  {Marcy}, {Wright}, {Isaacson}, {Boyajian}, {von Braun}, {Hillenbrand},
  {Hinkley}, {Carpenter}, \& {Brewer}}]{2012ApJ...751...97C}
{Crepp}, J.~R., {Johnson}, J.~A., {Fischer}, D.~A., {et~al.} 2012, \apj, 751,
  97

\bibitem[{{Cutri} \& {et al.}(2013)}]{2013yCat.2328....0C}
{Cutri}, R.~M. \& {et al.} 2013, VizieR Online Data Catalog, II/328

\bibitem[{{Davies} {et~al.}(2018){Davies}, {Esposito}, {Schmid}, {Taylor},
  {Agapito}, {Agudo Berbel}, {Baruffolo}, {Biliotti}, {Biller}, {Black},
  {Boehle}, {Briguglio}, {Buron}, {Carbonaro}, {Cortes}, {Cresci},
  {Deysenroth}, {Di Cianno}, {Di Rico}, {Doelman}, {Dolci}, {Dorn},
  {Eisenhauer}, {Fantinel}, {Ferruzzi}, {Feuchtgruber}, {F{\"o}rster
  Schreiber}, {Gao}, {Gemperlein}, {Genzel}, {George}, {Gillessen}, {Giordano},
  {Glauser}, {Glindemann}, {Grani}, {Hartl}, {Heijmans}, {Henry}, {Huber},
  {Kasper}, {Keller}, {Kenworthy}, {K{\"u}hn}, {Kuntschner}, {Lightfoot},
  {Lunney}, {MacIntosh}, {Mannucci}, {March}, {Neeser}, {Patapis}, {Pearson},
  {Plattner}, {Puglisi}, {Quanz}, {Rau}, {Riccardi}, {Salasnich}, {Schubert},
  {Snik}, {Sturm}, {Valentini}, {Waring}, {Wiezorrek}, \&
  {Xompero}}]{2018SPIE10702E..09D}
{Davies}, R., {Esposito}, S., {Schmid}, H.~M., {et~al.} 2018, in Society of
  Photo-Optical Instrumentation Engineers (SPIE) Conference Series, Vol. 10702,
  Ground-based and Airborne Instrumentation for Astronomy VII, 1070209

\bibitem[{{Dupuy} {et~al.}(2019){Dupuy}, {Brandt}, {Kratter}, \&
  {Bowler}}]{2019ApJ...871L...4D}
{Dupuy}, T.~J., {Brandt}, T.~D., {Kratter}, K.~M., \& {Bowler}, B.~P. 2019,
  \apj, 871, L4

\bibitem[{Dupuy {et~al.}(2019)Dupuy, Brandt, Kratter, \& Bowler}]{Dupuy_2019}
Dupuy, T.~J., Brandt, T.~D., Kratter, K.~M., \& Bowler, B.~P. 2019, The
  Astrophysical Journal, 871, L4

\bibitem[{{Feng} {et~al.}(2017){Feng}, {Tuomi}, {Jones}, {Barnes},
  {Anglada-Escud{\'e}}, {Vogt}, \& {Butler}}]{2017AJ....154..135F}
{Feng}, F., {Tuomi}, M., {Jones}, H.~R.~A., {et~al.} 2017, \aj, 154, 135

\bibitem[{{Gaia Collaboration} {et~al.}(2018){Gaia Collaboration}, {Brown},
  {Vallenari}, {Prusti}, {de Bruijne}, {Babusiaux}, {Bailer-Jones}, {Biermann},
  {Evans}, {Eyer}, {Jansen}, {Jordi}, {Klioner}, {Lammers}, {Lindegren},
  {Luri}, {Mignard}, {Panem}, {Pourbaix}, {Randich}, {Sartoretti}, {Siddiqui},
  {Soubiran}, {van Leeuwen}, {Walton}, {Arenou}, {Bastian}, {Cropper},
  {Drimmel}, {Katz}, {Lattanzi}, {Bakker}, {Cacciari}, {Casta{\~n}eda},
  {Chaoul}, {Cheek}, {De Angeli}, {Fabricius}, {Guerra}, {Holl}, {Masana},
  {Messineo}, {Mowlavi}, {Nienartowicz}, {Panuzzo}, {Portell}, {Riello},
  {Seabroke}, {Tanga}, {Th{\'e}venin}, {Gracia-Abril}, {Comoretto},
  {Garcia-Reinaldos}, {Teyssier}, {Altmann}, {Andrae}, {Audard},
  {Bellas-Velidis}, {Benson}, {Berthier}, {Blomme}, {Burgess}, {Busso},
  {Carry}, {Cellino}, {Clementini}, {Clotet}, {Creevey}, {Davidson}, {De
  Ridder}, {Delchambre}, {Dell'Oro}, {Ducourant}, {Fern{\'a}ndez-
  Hern{\'a}ndez}, {Fouesneau}, {Fr{\'e}mat}, {Galluccio}, {Garc{\'\i}a-Torres},
  {Gonz{\'a}lez-N{\'u}{\~n}ez}, {Gonz{\'a}lez-Vidal}, {Gosset}, {Guy},
  {Halbwachs}, {Hambly}, {Harrison}, {Hern{\'a}ndez}, {Hestroffer}, {Hodgkin},
  {Hutton}, {Jasniewicz}, {Jean-Antoine-Piccolo}, {Jordan}, {Korn},
  {Krone-Martins}, {Lanzafame}, {Lebzelter}, {L{\"o}ffler}, {Manteiga},
  {Marrese}, {Mart{\'\i}n-Fleitas}, {Moitinho}, {Mora}, {Muinonen}, {Osinde},
  {Pancino}, {Pauwels}, {Petit}, {Recio-Blanco}, {Richards}, {Rimoldini},
  {Robin}, {Sarro}, {Siopis}, {Smith}, {Sozzetti}, {S{\"u}veges}, {Torra}, {van
  Reeven}, {Abbas}, {Abreu Aramburu}, {Accart}, {Aerts}, {Altavilla},
  {{\'A}lvarez}, {Alvarez}, {Alves}, {Anderson}, {Andrei}, {Anglada Varela},
  {Antiche}, {Antoja}, {Arcay}, {Astraatmadja}, {Bach}, {Baker},
  {Balaguer-N{\'u}{\~n}ez}, {Balm}, {Barache}, {Barata}, {Barbato}, {Barblan},
  {Barklem}, {Barrado}, {Barros}, {Barstow}, {Bartholom{\'e} Mu{\~n}oz},
  {Bassilana}, {Becciani}, {Bellazzini}, {Berihuete}, {Bertone}, {Bianchi},
  {Bienaym{\'e}}, {Blanco-Cuaresma}, {Boch}, {Boeche}, {Bombrun}, {Borrachero},
  {Bossini}, {Bouquillon}, {Bourda}, {Bragaglia}, {Bramante}, {Breddels},
  {Bressan}, {Brouillet}, {Br{\"u}semeister}, {Brugaletta}, {Bucciarelli},
  {Burlacu}, {Busonero}, {Butkevich}, {Buzzi}, {Caffau}, {Cancelliere},
  {Cannizzaro}, {Cantat-Gaudin}, {Carballo}, {Carlucci}, {Carrasco},
  {Casamiquela}, {Castellani}, {Castro-Ginard}, {Charlot}, {Chemin},
  {Chiavassa}, {Cocozza}, {Costigan}, {Cowell}, {Crifo}, {Crosta}, {Crowley},
  {Cuypers}, {Dafonte}, {Damerdji}, {Dapergolas}, {David}, {David}, {de
  Laverny}, {De Luise}, {De March}, {de Martino}, {de Souza}, {de Torres},
  {Debosscher}, {del Pozo}, {Delbo}, {Delgado}, {Delgado}, {Di Matteo},
  {Diakite}, {Diener}, {Distefano}, {Dolding}, {Drazinos}, {Dur{\'a}n},
  {Edvardsson}, {Enke}, {Eriksson}, {Esquej}, {Eynard Bontemps}, {Fabre},
  {Fabrizio}, {Faigler}, {Falc{\~a}o}, {Farr{\`a}s Casas}, {Federici},
  {Fedorets}, {Fernique}, {Figueras}, {Filippi}, {Findeisen}, {Fonti},
  {Fraile}, {Fraser}, {Fr{\'e}zouls}, {Gai}, {Galleti}, {Garabato},
  {Garc{\'\i}a-Sedano}, {Garofalo}, {Garralda}, {Gavel}, {Gavras}, {Gerssen},
  {Geyer}, {Giacobbe}, {Gilmore}, {Girona}, {Giuffrida}, {Glass}, {Gomes},
  {Granvik}, {Gueguen}, {Guerrier}, {Guiraud}, {Guti{\'e}rrez-S{\'a}nchez},
  {Haigron}, {Hatzidimitriou}, {Hauser}, {Haywood}, {Heiter}, {Helmi}, {Heu},
  {Hilger}, {Hobbs}, {Hofmann}, {Holland}, {Huckle}, {Hypki}, {Icardi},
  {Jan{\ss}en}, {Jevardat de Fombelle}, {Jonker}, {Juh{\'a}sz}, {Julbe},
  {Karampelas}, {Kewley}, {Klar}, {Kochoska}, {Kohley}, {Kolenberg},
  {Kontizas}, {Kontizas}, {Koposov}, {Kordopatis}, {Kostrzewa-Rutkowska},
  {Koubsky}, {Lambert}, {Lanza}, {Lasne}, {Lavigne}, {Le Fustec}, {Le
  Poncin-Lafitte}, {Lebreton}, {Leccia}, {Leclerc}, {Lecoeur-Taibi},
  {Lenhardt}, {Leroux}, {Liao}, {Licata}, {Lindstr{\o}m}, {Lister}, {Livanou},
  {Lobel}, {L{\'o}pez}, {Managau}, {Mann}, {Mantelet}, {Marchal}, {Marchant},
  {Marconi}, {Marinoni}, {Marschalk{\'o}}, {Marshall}, {Martino}, {Marton},
  {Mary}, {Massari}, {Matijevi{\v{c}}}, {Mazeh}, {McMillan}, {Messina},
  {Michalik}, {Millar}, {Molina}, {Molinaro}, {Moln{\'a}r}, {Montegriffo},
  {Mor}, {Morbidelli}, {Morel}, {Morris}, {Mulone}, {Muraveva}, {Musella},
  {Nelemans}, {Nicastro}, {Noval}, {O'Mullane}, {Ord{\'e}novic},
  {Ord{\'o}{\~n}ez-Blanco}, {Osborne}, {Pagani}, {Pagano}, {Pailler},
  {Palacin}, {Palaversa}, {Panahi}, {Pawlak}, {Piersimoni}, {Pineau}, {Plachy},
  {Plum}, {Poggio}, {Poujoulet}, {Pr{\v{s}}a}, {Pulone}, {Racero}, {Ragaini},
  {Rambaux}, {Ramos-Lerate}, {Regibo}, {Reyl{\'e}}, {Riclet}, {Ripepi}, {Riva},
  {Rivard}, {Rixon}, {Roegiers}, {Roelens}, {Romero-G{\'o}mez}, {Rowell},
  {Royer}, {Ruiz-Dern}, {Sadowski}, {Sagrist{\`a} Sell{\'e}s}, {Sahlmann},
  {Salgado}, {Salguero}, {Sanna}, {Santana- Ros}, {Sarasso}, {Savietto},
  {Schultheis}, {Sciacca}, {Segol}, {Segovia}, {S{\'e}gransan}, {Shih},
  {Siltala}, {Silva}, {Smart}, {Smith}, {Solano}, {Solitro}, {Sordo}, {Soria
  Nieto}, {Souchay}, {Spagna}, {Spoto}, {Stampa}, {Steele},
  {Steidelm{\"u}ller}, {Stephenson}, {Stoev}, {Suess}, {Surdej}, {Szabados},
  {Szegedi-Elek}, {Tapiador}, {Taris}, {Tauran}, {Taylor}, {Teixeira},
  {Terrett}, {Teyssandier}, {Thuillot}, {Titarenko}, {Torra Clotet}, {Turon},
  {Ulla}, {Utrilla}, {Uzzi}, {Vaillant}, {Valentini}, {Valette}, {van Elteren},
  {Van Hemelryck}, {van Leeuwen}, {Vaschetto}, {Vecchiato}, {Veljanoski},
  {Viala}, {Vicente}, {Vogt}, {von Essen}, {Voss}, {Votruba}, {Voutsinas},
  {Walmsley}, {Weiler}, {Wertz}, {Wevers}, {Wyrzykowski}, {Yoldas},
  {{\v{Z}}erjal}, {Ziaeepour}, {Zorec}, {Zschocke}, {Zucker}, {Zurbach}, \&
  {Zwitter}}]{2018A&A...616A...1G}
{Gaia Collaboration}, {Brown}, A.~G.~A., {Vallenari}, A., {et~al.} 2018, \aap,
  616, A1

\bibitem[{{G{\'a}sp{\'a}r} {et~al.}(2013){G{\'a}sp{\'a}r}, {Rieke}, \&
  {Balog}}]{2013ApJ...768...25G}
{G{\'a}sp{\'a}r}, A., {Rieke}, G.~H., \& {Balog}, Z. 2013, \apj, 768, 25

\bibitem[{{Heinze} {et~al.}(2010){Heinze}, {Hinz}, {Sivanandam}, {Kenworthy},
  {Meyer}, \& {Miller}}]{2010ApJ...714.1551H}
{Heinze}, A.~N., {Hinz}, P.~M., {Sivanandam}, S., {et~al.} 2010, \apj, 714,
  1551

\bibitem[{{Janson}(2010)}]{2010MNRAS.408..514J}
{Janson}, M. 2010, \mnras, 408, 514

\bibitem[{{Jensen-Clem} {et~al.}(2018){Jensen-Clem}, {Mawet}, {Gomez Gonzalez},
  {Absil}, {Belikov}, {Currie}, {Kenworthy}, {Marois}, {Mazoyer}, {Ruane},
  {Tanner}, \& {Cantalloube}}]{2018AJ....155...19J}
{Jensen-Clem}, R., {Mawet}, D., {Gomez Gonzalez}, C.~A., {et~al.} 2018, \aj,
  155, 19

\bibitem[{{Johnson} {et~al.}(2010){Johnson}, {Aller}, {Howard}, \&
  {Crepp}}]{2010PASP..122..905J}
{Johnson}, J.~A., {Aller}, K.~M., {Howard}, A.~W., \& {Crepp}, J.~R. 2010,
  Publications of the Astronomical Society of the Pacific, 122, 905

\bibitem[{{Kammerer} \& {Quanz}(2018)}]{2018A&A...609A...4K}
{Kammerer}, J. \& {Quanz}, S.~P. 2018, \aap, 609, A4

\bibitem[{{Keenan} \& {McNeil}(1989)}]{1989ApJS...71..245K}
{Keenan}, P.~C. \& {McNeil}, R.~C. 1989, The Astrophysical Journal Supplement
  Series, 71, 245

\bibitem[{{Kenworthy} {et~al.}(2018){Kenworthy}, {Snik}, {Keller}, {Doelman},
  {Por}, {Absil}, {Carlomagno}, {Karlsson}, {Huby}, {Glauser}, {Quanz}, \&
  {Taylor}}]{2018SPIE10702E..46K}
{Kenworthy}, M.~A., {Snik}, F., {Keller}, C.~U., {et~al.} 2018, in Society of
  Photo-Optical Instrumentation Engineers (SPIE) Conference Series, Vol. 10702,
  1070246

\bibitem[{{Kervella} {et~al.}(2018){Kervella}, {Arenou}, {Mignard}, \&
  {Th{\'e}venin}}]{2018arXiv181108902K}
{Kervella}, P., {Arenou}, F., {Mignard}, F., \& {Th{\'e}venin}, F. 2018, arXiv
  e-prints, arXiv:1811.08902

\bibitem[{{Kidger} \& {Mart{\'\i}n-Luis}(2003)}]{2003AJ....125.3311K}
{Kidger}, M.~R. \& {Mart{\'\i}n-Luis}, F. 2003, \aj, 125, 3311

\bibitem[{{Kirkpatrick} {et~al.}(1991){Kirkpatrick}, {Henry}, \&
  {McCarthy}}]{1991ApJS...77..417K}
{Kirkpatrick}, J.~D., {Henry}, T.~J., \& {McCarthy}, Donald~W., J. 1991, The
  Astrophysical Journal Supplement Series, 77, 417

\bibitem[{{Lasker} {et~al.}(2008){Lasker}, {Lattanzi}, {McLean}, {Bucciarelli},
  {Drimmel}, {Garcia}, {Greene}, {Guglielmetti}, {Hanley}, {Hawkins},
  {Laidler}, {Loomis}, {Meakes}, {Mignani}, {Morbidelli}, {Morrison},
  {Pannunzio}, {Rosenberg}, {Sarasso}, {Smart}, {Spagna}, {Sturch},
  {Volpicelli}, {White}, {Wolfe}, \& {Zacchei}}]{2008AJ....136..735L}
{Lasker}, B.~M., {Lattanzi}, M.~G., {McLean}, B.~J., {et~al.} 2008, \aj, 136,
  735

\bibitem[{{Lawler} {et~al.}(2014){Lawler}, {Di Francesco}, {Kennedy},
  {Sibthorpe}, {Booth}, {Vandenbussche}, {Matthews}, {Holland}, {Greaves},
  {Wilner}, {Tuomi}, {Blommaert}, {de Vries}, {Dominik}, {Fridlund}, {Gear},
  {Heras}, {Ivison}, \& {Olofsson}}]{2014MNRAS.444.2665L}
{Lawler}, S.~M., {Di Francesco}, J., {Kennedy}, G.~M., {et~al.} 2014, \mnras,
  444, 2665

\bibitem[{{Lenzen} {et~al.}(2003){Lenzen}, {Hartung}, {Brandner}, {Finger},
  {Hubin}, {Lacombe}, {Lagrange}, {Lehnert}, {Moorwood}, \&
  {Mouillet}}]{2003SPIE.4841..944L}
{Lenzen}, R., {Hartung}, M., {Brandner}, W., {et~al.} 2003, in \procspie, Vol.
  4841, Instrument Design and Performance for Optical/Infrared Ground-based
  Telescopes, ed. M.~{Iye} \& A.~F.~M. {Moorwood}, 944--952

\bibitem[{{Lindegren} {et~al.}(2018){Lindegren}, {Hern{\'a}ndez}, {Bombrun},
  {Klioner}, {Bastian}, {Ramos-Lerate}, {de Torres}, {Steidelm{\"u}ller},
  {Stephenson}, {Hobbs}, {Lammers}, {Biermann}, {Geyer}, {Hilger}, {Michalik},
  {Stampa}, {McMillan}, {Casta{\~n}eda}, {Clotet}, {Comoretto}, {Davidson},
  {Fabricius}, {Gracia}, {Hambly}, {Hutton}, {Mora}, {Portell}, {van Leeuwen},
  {Abbas}, {Abreu}, {Altmann}, {Andrei}, {Anglada}, {Balaguer-N{\'u}{\~n}ez},
  {Barache}, {Becciani}, {Bertone}, {Bianchi}, {Bouquillon}, {Bourda},
  {Br{\"u}semeister}, {Bucciarelli}, {Busonero}, {Buzzi}, {Cancelliere},
  {Carlucci}, {Charlot}, {Cheek}, {Crosta}, {Crowley}, {de Bruijne}, {de
  Felice}, {Drimmel}, {Esquej}, {Fienga}, {Fraile}, {Gai}, {Garralda},
  {Gonz{\'a}lez- Vidal}, {Guerra}, {Hauser}, {Hofmann}, {Holl}, {Jordan},
  {Lattanzi}, {Lenhardt}, {Liao}, {Licata}, {Lister}, {L{\"o}ffler},
  {Marchant}, {Martin-Fleitas}, {Messineo}, {Mignard}, {Morbidelli}, {Poggio},
  {Riva}, {Rowell}, {Salguero}, {Sarasso}, {Sciacca}, {Siddiqui}, {Smart},
  {Spagna}, {Steele}, {Taris}, {Torra}, {van Elteren}, {van Reeven}, \&
  {Vecchiato}}]{2018A&A...616A...2L}
{Lindegren}, L., {Hern{\'a}ndez}, J., {Bombrun}, A., {et~al.} 2018, \aap, 616,
  A2

\bibitem[{{Linder} {et~al.}(2019){Linder}, {Mordasini}, {Molli{\`e}re},
  {Marleau}, {Malik}, {Quanz}, \& {Meyer}}]{2019A&A...623A..85L}
{Linder}, E.~F., {Mordasini}, C., {Molli{\`e}re}, P., {et~al.} 2019, \aap, 623,
  A85

\bibitem[{{Lo Curto} {et~al.}(2015){Lo Curto}, {Pepe}, {Avila}, {Boffin},
  {Bovay}, {Chazelas}, {Coffinet}, {Fleury}, {Hughes}, {Lovis}, {Maire},
  {Manescau}, {Pasquini}, {Rihs}, {Sinclaire}, \& {Udry}}]{2015Msngr.162....9L}
{Lo Curto}, G., {Pepe}, F., {Avila}, G., {et~al.} 2015, The Messenger, 162, 9

\bibitem[{{Lomb}(1976)}]{1976Ap&SS..39..447L}
{Lomb}, N.~R. 1976, \apss, 39, 447

\bibitem[{{Ma} {et~al.}(2018){Ma}, {Ge}, {Muterspaugh}, {Singer}, {Henry},
  {Gonz{\'a}lez Hern{\'a}ndez}, {Sithajan}, {Jeram}, {Williamson}, {Stassun},
  {Kimock}, {Varosi}, {Schofield}, {Liu}, {Powell}, {Cassette}, {Jakeman},
  {Avner}, {Grieves}, {Barnes}, {Zhao}, {Gilda}, {Grantham}, {Stafford},
  {Savage}, {Bland}, \& {Ealey}}]{2018MNRAS.480.2411M}
{Ma}, B., {Ge}, J., {Muterspaugh}, M., {et~al.} 2018, \mnras, 480, 2411

\bibitem[{{Macintosh} {et~al.}(2014){Macintosh}, {Graham}, {Ingraham},
  {Konopacky}, {Marois}, {Perrin}, {Poyneer}, {Bauman}, {Barman}, {Burrows},
  {Cardwell}, {Chilcote}, {De Rosa}, {Dillon}, {Doyon}, {Dunn}, {Erikson},
  {Fitzgerald}, {Gavel}, {Goodsell}, {Hartung}, {Hibon}, {Kalas}, {Larkin},
  {Maire}, {Marchis}, {Marley}, {McBride}, {Millar- Blanchaer}, {Morzinski},
  {Norton}, {Oppenheimer}, {Palmer}, {Patience}, {Pueyo}, {Rantakyro},
  {Sadakuni}, {Saddlemyer}, {Savransky}, {Serio}, {Soummer},
  {Sivaramakrishnan}, {Song}, {Thomas}, {Wallace}, {Wiktorowicz}, \&
  {Wolff}}]{2014PNAS..11112661M}
{Macintosh}, B., {Graham}, J.~R., {Ingraham}, P., {et~al.} 2014, Proceedings of
  the National Academy of Science, 111, 12661

\bibitem[{{Mamajek} \& {Hillenbrand}(2008)}]{2008ApJ...687.1264M}
{Mamajek}, E.~E. \& {Hillenbrand}, L.~A. 2008, \apj, 687, 1264

\bibitem[{{Marconi} {et~al.}(2018){Marconi}, {Allende Prieto}, {Amado},
  {Amate}, {Augusto}, {Becerril}, {Bezawada}, {Boisse}, {Bouchy}, {Cabral},
  {Chazelas}, {Cirami}, {Coretti}, {Cristiani}, {Cupani}, {de Castro Le{\~a}o},
  {de Medeiros}, {de Souza}, {Di Marcantonio}, {Di Varano}, {D'Odorico},
  {Drass}, {Figueira}, {Fragoso}, {Fynbo}, {Genoni}, {Gonz{\'a}lez
  Hern{\'a}ndez}, {Haehnelt}, {Hughes}, {Huke}, {Kjeldsen}, {Korn}, {Land oni},
  {Liske}, {Lovis}, {Maiolino}, {Marquart}, {Martins}, {Mason}, {Monteiro},
  {Morris}, {Murray}, {Niedzielski}, {Oliva}, {Origlia}, {Pall{\'e}},
  {Parr-Burman}, {Parro}, {Pepe}, {Piskunov}, {Rasilla}, {Rees}, {Rebolo},
  {Riva}, {Rousseau}, {Sanna}, {Santos}, {Shen}, {Sortino}, {Sosnowska},
  {Sousa}, {Stempels}, {Strassmeier}, {Tenegi}, {Tozzi}, {Udry}, {Valenziano},
  {Vanzi}, {Weber}, {Woche}, {Xompero}, \& {Zackrisson}}]{2018SPIE10702E..1YM}
{Marconi}, A., {Allende Prieto}, C., {Amado}, P.~J., {et~al.} 2018, in Society
  of Photo-Optical Instrumentation Engineers (SPIE) Conference Series, Vol.
  10702, 107021Y

\bibitem[{{Marley} {et~al.}(2007){Marley}, {Fortney}, {Hubickyj},
  {Bodenheimer}, \& {Lissauer}}]{2007ApJ...655..541M}
{Marley}, M.~S., {Fortney}, J.~J., {Hubickyj}, O., {Bodenheimer}, P., \&
  {Lissauer}, J.~J. 2007, \apj, 655, 541

\bibitem[{{Marois} {et~al.}(2006){Marois}, {Lafreni{\`e}re}, {Doyon},
  {Macintosh}, \& {Nadeau}}]{2006ApJ...641..556M}
{Marois}, C., {Lafreni{\`e}re}, D., {Doyon}, R., {Macintosh}, B., \& {Nadeau},
  D. 2006, \apj, 641, 556

\bibitem[{Mawet {et~al.}(2019)Mawet, Hirsch, Lee, Ruffio, Bottom, Fulton,
  Absil, Beichman, Bowler, Bryan, Choquet, Ciardi, Christiaens, Defr{\`{e}}re,
  Gonzalez, Howard, Huby, Isaacson, Jensen-Clem, Kosiarek, Marcy, Meshkat,
  Petigura, Reggiani, Ruane, Serabyn, Sinukoff, Wang, Weiss, \&
  Ygouf}]{Mawet_2019}
Mawet, D., Hirsch, L., Lee, E.~J., {et~al.} 2019, The Astronomical Journal,
  157, 33

\bibitem[{{Mawet} {et~al.}(2014){Mawet}, {Milli}, {Wahhaj}, {Pelat}, {Absil},
  {Delacroix}, {Boccaletti}, {Kasper}, {Kenworthy}, {Marois}, {Mennesson}, \&
  {Pueyo}}]{2014ApJ...792...97M}
{Mawet}, D., {Milli}, J., {Wahhaj}, Z., {et~al.} 2014, \apj, 792, 97

\bibitem[{{Mayor} {et~al.}(2003){Mayor}, {Pepe}, {Queloz}, {Bouchy},
  {Rupprecht}, {Lo Curto}, {Avila}, {Benz}, {Bertaux}, {Bonfils}, {Dall},
  {Dekker}, {Delabre}, {Eckert}, {Fleury}, {Gilliotte}, {Gojak}, {Guzman},
  {Kohler}, {Lizon}, {Longinotti}, {Lovis}, {Megevand}, {Pasquini}, {Reyes},
  {Sivan}, {Sosnowska}, {Soto}, {Udry}, {van Kesteren}, {Weber}, \&
  {Weilenmann}}]{2003Msngr.114...20M}
{Mayor}, M., {Pepe}, F., {Queloz}, D., {et~al.} 2003, The Messenger, 114, 20

\bibitem[{{Mennesson} {et~al.}(2016){Mennesson}, {Gaudi}, {Seager}, {Cahoy},
  {Domagal-Goldman}, {Feinberg}, {Guyon}, {Kasdin}, {Marois}, {Mawet},
  {Tamura}, {Mouillet}, {Prusti}, {Quirrenbach}, {Robinson}, {Rogers},
  {Scowen}, {Somerville}, {Stapelfeldt}, {Stern}, {Still}, {Turnbull}, {Booth},
  {Kiessling}, {Kuan}, \& {Warfield}}]{2016SPIE.9904E..0LM}
{Mennesson}, B., {Gaudi}, S., {Seager}, S., {et~al.} 2016, in Society of
  Photo-Optical Instrumentation Engineers (SPIE) Conference Series, Vol. 9904,
  Space Telescopes and Instrumentation 2016: Optical, Infrared, and Millimeter
  Wave, 99040L

\bibitem[{{Meyer} {et~al.}(2018){Meyer}, {Amara}, {Reggiani}, \&
  {Quanz}}]{2018A&A...612L...3M}
{Meyer}, M.~R., {Amara}, A., {Reggiani}, M., \& {Quanz}, S.~P. 2018, \aap, 612,
  L3

\bibitem[{{Morel} \& {Magnenat}(1978)}]{1978A&AS...34..477M}
{Morel}, M. \& {Magnenat}, P. 1978, Astronomy and Astrophysics Supplement
  Series, 34, 477

\bibitem[{{Nakajima} {et~al.}(1995){Nakajima}, {Oppenheimer}, {Kulkarni},
  {Golimowski}, {Matthews}, \& {Durrance}}]{1995Natur.378..463N}
{Nakajima}, T., {Oppenheimer}, B.~R., {Kulkarni}, S.~R., {et~al.} 1995, \nat,
  378, 463

\bibitem[{{Nakajima} {et~al.}(2015){Nakajima}, {Tsuji}, \&
  {Takeda}}]{2015AJ....150...53N}
{Nakajima}, T., {Tsuji}, T., \& {Takeda}, Y. 2015, \aj, 150, 53

\bibitem[{{Oppenheimer} {et~al.}(1995){Oppenheimer}, {Kulkarni}, {Matthews}, \&
  {Nakajima}}]{1995Sci...270.1478O}
{Oppenheimer}, B.~R., {Kulkarni}, S.~R., {Matthews}, K., \& {Nakajima}, T.
  1995, Science, 270, 1478

\bibitem[{{Pepe} {et~al.}(2002){Pepe}, {Mayor}, {Rupprecht}, {Avila},
  {Ballester}, {Beckers}, {Benz}, {Bertaux}, {Bouchy}, {Buzzoni}, {Cavadore},
  {Deiries}, {Dekker}, {Delabre}, {D'Odorico}, {Eckert}, {Fischer}, {Fleury},
  {George}, {Gilliotte}, {Gojak}, {Guzman}, {Koch}, {Kohler}, {Kotzlowski},
  {Lacroix}, {Le Merrer}, {Lizon}, {Lo Curto}, {Longinotti}, {Megevand},
  {Pasquini}, {Petitpas}, {Pichard}, {Queloz}, {Reyes}, {Richaud}, {Sivan},
  {Sosnowska}, {Soto}, {Udry}, {Ureta}, {van Kesteren}, {Weber}, {Weilenmann},
  {Wicenec}, {Wieland}, {Christensen-Dalsgaard}, {Dravins}, {Hatzes},
  {K{\"u}rster}, {Paresce}, \& {Penny}}]{2002Msngr.110....9P}
{Pepe}, F., {Mayor}, M., {Rupprecht}, G., {et~al.} 2002, The Messenger, 110, 9

\bibitem[{{Perryman} {et~al.}(2014){Perryman}, {Hartman}, {Bakos}, \&
  {Lindegren}}]{2014ApJ...797...14P}
{Perryman}, M., {Hartman}, J., {Bakos}, G.~{\'A}., \& {Lindegren}, L. 2014,
  \apj, 797, 14

\bibitem[{{Pueyo} {et~al.}(2017){Pueyo}, {Zimmerman}, {Bolcar}, {Groff},
  {Stark}, {Ruane}, {Jewell}, {Soummer}, {St. Laurent}, {Wang}, {Redding},
  {Mazoyer}, {Fogarty}, {Juanola-Parramon}, {Domagal-Goldman}, {Roberge},
  {Guyon}, \& {Mandell}}]{2017SPIE10398E..0FP}
{Pueyo}, L., {Zimmerman}, N., {Bolcar}, M., {et~al.} 2017, in Society of
  Photo-Optical Instrumentation Engineers (SPIE) Conference Series, Vol. 10398,
  103980F

\bibitem[{{Quanz} {et~al.}(2012){Quanz}, {Crepp}, {Janson}, {Avenhaus},
  {Meyer}, \& {Hillenbrand}}]{2012ApJ...754..127Q}
{Quanz}, S.~P., {Crepp}, J.~R., {Janson}, M., {et~al.} 2012, \apj, 754, 127

\bibitem[{{Quanz} {et~al.}(2015){Quanz}, {Crossfield}, {Meyer}, {Schmalzl}, \&
  {Held}}]{2015IJAsB..14..279Q}
{Quanz}, S.~P., {Crossfield}, I., {Meyer}, M.~R., {Schmalzl}, E., \& {Held}, J.
  2015, International Journal of Astrobiology, 14, 279

\bibitem[{{Ranalli} {et~al.}(2018){Ranalli}, {Hobbs}, \&
  {Lindegren}}]{2018A&A...614A..30R}
{Ranalli}, P., {Hobbs}, D., \& {Lindegren}, L. 2018, \aap, 614, A30

\bibitem[{{Ricker} {et~al.}(2015){Ricker}, {Winn}, {Vanderspek}, {Latham},
  {Bakos}, {Bean}, {Berta-Thompson}, {Brown}, {Buchhave}, {Butler}, {Butler},
  {Chaplin}, {Charbonneau}, {Christensen-Dalsgaard}, {Clampin}, {Deming},
  {Doty}, {De Lee}, {Dressing}, {Dunham}, {Endl}, {Fressin}, {Ge}, {Henning},
  {Holman}, {Howard}, {Ida}, {Jenkins}, {Jernigan}, {Johnson}, {Kaltenegger},
  {Kawai}, {Kjeldsen}, {Laughlin}, {Levine}, {Lin}, {Lissauer}, {MacQueen},
  {Marcy}, {McCullough}, {Morton}, {Narita}, {Paegert}, {Palle}, {Pepe},
  {Pepper}, {Quirrenbach}, {Rinehart}, {Sasselov}, {Sato}, {Seager},
  {Sozzetti}, {Stassun}, {Sullivan}, {Szentgyorgyi}, {Torres}, {Udry}, \&
  {Villasenor}}]{2015JATIS...1a4003R}
{Ricker}, G.~R., {Winn}, J.~N., {Vanderspek}, R., {et~al.} 2015, Journal of
  Astronomical Telescopes, Instruments, and Systems, 1, 014003

\bibitem[{{Robertson} {et~al.}(2015){Robertson}, {Roy}, \&
  {Mahadevan}}]{2015ApJ...805L..22R}
{Robertson}, P., {Roy}, A., \& {Mahadevan}, S. 2015, \apjl, 805, L22

\bibitem[{{Rousset} {et~al.}(2003){Rousset}, {Lacombe}, {Puget}, {Hubin},
  {Gendron}, {Fusco}, {Arsenault}, {Charton}, {Feautrier}, {Gigan}, {Kern},
  {Lagrange}, {Madec}, {Mouillet}, {Rabaud}, {Rabou}, {Stadler}, \&
  {Zins}}]{2003SPIE.4839..140R}
{Rousset}, G., {Lacombe}, F., {Puget}, P., {et~al.} 2003, in \procspie, Vol.
  4839, Adaptive Optical System Technologies II, ed. P.~L. {Wizinowich} \&
  D.~{Bonaccini}, 140--149

\bibitem[{{Scargle}(1982)}]{1982ApJ...263..835S}
{Scargle}, J.~D. 1982, \apj, 263, 835

\bibitem[{{Snellen} \& {Brown}(2018)}]{2018NatAs...2..883S}
{Snellen}, I.~A.~G. \& {Brown}, A.~G.~A. 2018, Nature Astronomy, 2, 883

\bibitem[{{Sousa} {et~al.}(2008){Sousa}, {Santos}, {Mayor}, {Udry},
  {Casagrande}, {Israelian}, {Pepe}, {Queloz}, \&
  {Monteiro}}]{2008A&A...487..373S}
{Sousa}, S.~G., {Santos}, N.~C., {Mayor}, M., {et~al.} 2008, \aap, 487, 373

\bibitem[{{Stolker} {et~al.}(2019){Stolker}, {Bonse}, {Quanz}, {Amara},
  {Cugno}, {Bohn}, \& {Boehle}}]{2019A&A...621A..59S}
{Stolker}, T., {Bonse}, M.~J., {Quanz}, S.~P., {et~al.} 2019, \aap, 621, A59

\bibitem[{{Takeda} {et~al.}(2007){Takeda}, {Kawanomoto}, {Honda}, {Ando}, \&
  {Sakurai}}]{2007A&A...468..663T}
{Takeda}, Y., {Kawanomoto}, S., {Honda}, S., {Ando}, H., \& {Sakurai}, T. 2007,
  \aap, 468, 663

\bibitem[{Thalmann {et~al.}(2011)Thalmann, Usuda, Kenworthy, Janson, Mamajek,
  Brandner, Dominik, Goto, Hayano, Henning, Hinz, Minowa, \&
  Tamura}]{Thalmann_2011}
Thalmann, C., Usuda, T., Kenworthy, M., {et~al.} 2011, The Astrophysical
  Journal, 732, L34

\bibitem[{{Tuomi} {et~al.}(2014){Tuomi}, {Jones}, {Barnes},
  {Anglada-Escud{\'e}}, \& {Jenkins}}]{2014MNRAS.441.1545T}
{Tuomi}, M., {Jones}, H. R.~A., {Barnes}, J.~R., {Anglada-Escud{\'e}}, G., \&
  {Jenkins}, J.~S. 2014, \mnras, 441, 1545

\bibitem[{{Tuomi} {et~al.}(2013){Tuomi}, {Jones}, {Jenkins}, {Tinney},
  {Butler}, {Vogt}, {Barnes}, {Wittenmyer}, {O'Toole}, {Horner}, {Bailey},
  {Carter}, {Wright}, {Salter}, \& {Pinfield}}]{2013A&A...551A..79T}
{Tuomi}, M., {Jones}, H.~R.~A., {Jenkins}, J.~S., {et~al.} 2013, \aap, 551, A79

\bibitem[{{van Leeuwen}(2007)}]{2007A&A...474..653V}
{van Leeuwen}, F. 2007, \aap, 474, 653

\bibitem[{{Vican}(2012)}]{2012AJ....143..135V}
{Vican}, L. 2012, \aj, 143, 135

\bibitem[{{Vigan} {et~al.}(2015){Vigan}, {Gry}, {Salter}, {Mesa}, {Homeier},
  {Moutou}, \& {Allard}}]{2015MNRAS.454..129V}
{Vigan}, A., {Gry}, C., {Salter}, G., {et~al.} 2015, \mnras, 454, 129

\bibitem[{{Vogt} {et~al.}(1994){Vogt}, {Allen}, {Bigelow}, {Bresee}, {Brown},
  {Cantrall}, {Conrad}, {Couture}, {Delaney}, {Epps}, {Hilyard}, {Hilyard},
  {Horn}, {Jern}, {Kanto}, {Keane}, {Kibrick}, {Lewis}, {Osborne},
  {Pardeilhan}, {Pfister}, {Ricketts}, {Robinson}, {Stover}, {Tucker}, {Ward},
  \& {Wei}}]{1994SPIE.2198..362V}
{Vogt}, S.~S., {Allen}, S.~L., {Bigelow}, B.~C., {et~al.} 1994, in \procspie,
  Vol. 2198, Instrumentation in Astronomy VIII, ed. D.~L. {Crawford} \& E.~R.
  {Craine}, 362

\bibitem[{{Yee} {et~al.}(2017){Yee}, {Petigura}, \& {von
  Braun}}]{2017ApJ...836...77Y}
{Yee}, S.~W., {Petigura}, E.~A., \& {von Braun}, K. 2017, \apj, 836, 77

\end{thebibliography}

\end{document}